\documentclass{aa}  

\usepackage{graphicx}
\usepackage{natbib}
\usepackage{txfonts}
\newcommand{\mo}{$\mathrm{O_2}$}

\usepackage{natbib,twoopt}
\usepackage[breaklinks=true]{hyperref} %% to avoid \citeads line fills
\bibpunct{(}{)}{;}{a}{}{,}             %% natbib format for A&A and ApJ
\makeatletter
  \newcommandtwoopt{\citeads}[3][][]{\href{http://adsabs.harvard.edu/abs/#3}%
    {\def\hyper@linkstart##1##2{}%
     \let\hyper@linkend\@empty\citealp[#1][#2]{#3}}}
  \newcommandtwoopt{\citepads}[3][][]{\href{http://adsabs.harvard.edu/abs/#3}%
    {\def\hyper@linkstart##1##2{}%
     \let\hyper@linkend\@empty\citep[#1][#2]{#3}}}
  \newcommandtwoopt{\citetads}[3][][]{\href{http://adsabs.harvard.edu/abs/#3}%
    {\def\hyper@linkstart##1##2{}%
     \let\hyper@linkend\@empty\citet[#1][#2]{#3}}}
  \newcommandtwoopt{\citeyearads}[3][][]%
    {\href{http://adsabs.harvard.edu/abs/#3}
    {\def\hyper@linkstart##1##2{}%
     \let\hyper@linkend\@empty\citeyear[#1][#2]{#3}}}
\makeatother

\begin{document}

   \title{First on-sky results of a FIOS prototype, a Fabry Perot Based Instrument for Oxygen Searches}
   \titlerunning{FIOS prototype}

   \author{S. Rukdee
          \inst{1}
          \and
          S. Ben-Ami\inst{2}
          \and
          M. López-Morales\inst{3}
          \and 
          A. Szentgyorgyi\inst{3}
          \and \\
          D. Charbonneau\inst{3} 
          \and 
          J. Garcia-Mejia\inst{3}
          \and 
          J. Buchner\inst{1}
          }

   \institute{Max Planck Institute for Extraterrestrial     Physics, Giessenbachstrasse, 85748 Garching, Germany
         \and
         Department of Particle Physics and Astrophysics, Weizmann Institute of Science, Rehovot 7610001, Israel
         \and
         Center for Astrophysics ${\rm \mid}$ Harvard {\rm \&} Smithsonian, 60 Garden Street, Cambridge, MA 02138, USA \\
             }

   \date{Received 07 April 2023 / Accepted 24 July 2023}

% \abstract{}{}{}{}{} 
% 5 {} token are mandatory
 
  \abstract
  % context heading (optional) 
  % {} leave it empty if necessary  
    {The upcoming Extremely Large Telescopes (ELTs) are expected to have the collecting area required to detect potential biosignature gases such as molecular oxygen, $\mathrm{O_2}$, in the atmosphere of terrestrial planets around nearby stars.}
  % aims heading (mandatory)
   {One of the most promising detection methods is transmission spectroscopy. To maximize our capability to detect $\mathrm{O_2}$ using this method, spectral resolutions $\mathrm{R}\geq 300,000$ are required to fully resolve the absorption lines in an Earth-like exoplanet atmosphere and disentangle the signal from telluric lines.}
  % methods heading (mandatory)
   {Current high-resolution spectrographs typically achieve a spectral resolution of $\mathrm{R}\sim100,000$. Increasing the resolution in seeing limited observations/instruments requires drastically larger optical components, making these instruments even more expensive and hard to fabricate and assemble. Instead, we demonstrate a new approach to high-resolution spectroscopy. We implemented an ultra-high spectral resolution booster to be coupled in front of a high-resolution spectrograph. The instrument is based on a chained Fabry Perot array which generates a hyperfine spectral profile.}
  % results heading (mandatory)
   {We present on-sky telluric observations with a lab demonstrator. Depending on the configuration, this two-arm prototype reaches a resolution of R=250,000-350,000. After carefully modeling the prototype's behavior, we propose a Fabry Perot Interferometer (FPI) design for an eight-arm array configuration aimed at ELTs capable of exceeding R=300,000.}
  % conclusions heading (optional), leave it empty if necessary 
   {The novel FPI resolution booster can be plugged in at the front end of an existing R=100,000 spectrograph to overwrite the spectral profile with a higher resolution for exoplanet atmosphere studies.}


   \maketitle
%
%-------------------------------------------------------------------

\section{Introduction} \label{sec:intro}

Among the confirmed exoplanets detected in the past $\sim25\,$years, a couple of dozen exoplanets have a rocky composition, also known as terrestrial exoplanets. Missions like Kepler \citep{Borucki_2010} and TESS \citep{Ricker_TESS_2014} have shown that these planets are among the most common in the solar neighborhood \citep{Dressing_Charbonneau2015}. A natural next step in the study of terrestrial exoplanets is the characterization of their atmospheres, a subject that remains largely unconstrained \citep{Wordsworth_Kreidberg2022}. Of specific interest are habitable zone planets, whose orbit allows the existence of liquid water on their surface. Studying the atmospheres of these planets will allow us to place Earth in a larger context, and understand the prevalence of the conditions which allowed life to evolve on our own planet. The possible detection of biomarkers in such studies represents an exciting opportunity for the scientific community. 

The most common and prolific studies of exoplanet atmospheres are through transit spectroscopy, in which we study the chemical composition of exoplanet atmospheres using spectra obtained during transits (\citealt{2000ApJ...537..916S,Charbonneau_2002}). As a planet passes through the line of sight between the observer and the host star, atmospheric molecules leave spectral imprints through absorption lines. The geometric cross-section of an exoplanet atmosphere is minute when compared to the surface area of a star ($10^{-4}-10^{-6})$, high signal-to-noise ratio (SNR) observations are key for this method. Recent studies have unveiled several species in gas giant atmospheres using high-resolution transmission spectroscopy. For example, the detection of CO \citep{Snellen2010,de_Kok_2013,Rodler_2012,Brogi_2012}, $\rm H_2O$ \citep{Brogi_2014,2017AJ....153..138B},  and TiO \citep{Hoeijmakers_2015} via cross-correlation technique \citep{Snellen2010}, and Doppler tomography \citep{Watson_2019}. These studies suggest that high-resolution spectra ($R \geq100,000$) favor the unambiguous detection of molecules in an exoplanet atmosphere. For a fixed SNR per resolution element, a higher resolving power instrument can probe weaker and narrower spectral lines \citep{2005hris.conf....3B}.

One enticing molecule that is yet to be detected in an exoplanet atmosphere is molecular oxygen (\mo{}). \mo{} is a potential indicator of bioactivity in exoplanets when using our own Earth as a template \citep{Sagan1993}. Although it is important to interpret O$_2$ in its environmental context \citep{2018AsBio..18..630M}, this molecule remains a prominent biosignature  \citep{1965Natur.207..568L,1967Icar....7..149H}, and developing the instrumental capability for its detection is a key ingredient in the search for life beyond the solar system. The source of \mo{} on our own planet is oxygenic photosynthesis \citep{2017AsBio..17.1022M}. Going back to the origin of life on our own planet, the first single-celled organism, cyanobacteria, was able to convert carbon dioxide (CO$_2$) into \mo{} through photosynthesis \citep{holland_2006}. The Earth's atmosphere became enriched in \mo{} during the Archean ($2.5-4.0$ billion years ago) and likely led to the flora and fauna that followed \cite{}. 

Atomic oxygen bonds quickly with other molecules, and has to be replenished regularly to reach the high concentration of \mo{} we observe in our own atmosphere. As \mo{} is an abundant gas that is evenly mixed throughout the atmosphere \citep{2018AsBio..18..630M}, it is accessible for remote detection through methods such as transit spectroscopy. While the presence of \mo{} in exoplanet atmospheres can be the result of abiotic processes, a detection of \mo{} can be a strong indicator of biological activity when analyzed in its stellar and planetary context (\citealt{2014ApJ...785L..20W,2015AsBio..15..119L,2015ApJ...806..249G,2015ApJ...812..137H,2014ApJ...792...90D,2018AsBio..18..630M}).

It is unlikely that space-based missions will detect \mo{} in an earth-like exoplanet atmosphere within the next decade. The low instrumental sensitivity and resolution of JWST at short wavelengths will likely prohibit detection in the main \mo{} absorption bands at the B-bands (650 nm), A-band (765 nm), and the near infrared (NIR) (1270 nm), see \cite{Wunderlich2019AA}. An intriguing alternative route for detecting \mo{} at $6.4\mu m$ was recently suggested by \cite{Fauchez_2020}. Regardless, the prospect of detection with earth-based observations seems more plausible. 

As the next-generation Giant Segmented Mirror Telescopes (GSMTs) start coming online, we expect their large collecting areas to enable the high SNR observations required to detect molecular oxygen in terrestrial exoplanet atmospheres \citep{Snellen2013,Rodler_2014}. Several instruments with high dispersion capabilities such as  G-CLEF \citep{2016SPIE.9908E..22S}, GMTNIRS \citep{2016SPIE.9908E..21J}, METIS \citep{2014SPIE.9147E..21B}, and HIRES/ANDES \citep[][]{2018SPIE10702E..1YM}, could contribute to the study of exoplanet atmospheres and the possible detection of \mo{}. These instruments reach a maximum resolution of R=150,000. \cite{2019AJ_Lopez}  suggests that ultra-high resolution in the range of R=300,000-400,000 is optimal for \mo{} detection in terrestrial exoplanet atmospheres. This is set by line-of-site effects and the expected narrow line profile due to refraction effects by lower layers in an exoplanet atmosphere. Additionally, at R $\geq$ 300,000, the Earth's telluric can be separated from the exoplanet's atmospheric spectrum by leveraging the latter's relative velocity.

Seeing-limited high-resolution echelle spectrographs are limited in their capability to achieve a resolution significantly larger than $\sim100,000$ for $6.5-10\,$m apertures due to conservation of etendue. A large collecting area implies a larger collimated beam, dispersion grating, and instrument as a whole. This problem becomes exacerbated when considering the GSMTs, and the dependency of the resolution on the telescope diameter for such instruments \citep[$R\propto\frac{1}{D}$; \textit{e.g.,}][chapter 12.2 and 12.3]{Schroeder2000}. As a result, a new instrument approach is required. A significant effort is invested in boosting the spectral resolution. For example, the precision radial velocity (PRV) community has developed pupil or image slicers to boost spectral resolution \textit{e.g.,} G-CLEF \citep{2016SPIE.9908E..22S}, CARMENES \citep{quirrenbach2014}, PEPSI \cite{strassmeier2015}, FIDEOS \citep{vanzi2018}, TARdYS \citep{rukdee2019}. However, this solution is limited by the detector area available on the spectrograph focal plane. Another alternative is the use of externally dispersed spectrographs to achieve extremely high resolutions, see Dispersed Fixed Delay Interferometer \citep[DFDI][]{2002PASP..114.1016G,Ge_2006} and the TripleSpec Exoplanet Discovery Instrument \citep[TEDI][]{10.1117/12.735474} for Doppler exoplanet surveys. DFDI optimizes the optical delay in the interferometer and reduces photon noise by measuring multiple fringes over broadband. The interferometer is then coupled with a low- to medium-resolution post-disperser \citep{2002ApJ...571L.165G}. TEDI has an Externally Dispersed Interferometry (EDI) coupled with a conventional high-resolution R=20,000 spectrograph. The prototype demonstrated a factor of six increase in resolving power \citep{10.1117/12.735474}, but at the cost of exposure time, as at any given time, one samples only a limited subspace of the frequency range spanned by the signal in Fourier space.

In light of these results and challenges, \cite{Ben_Ami_2018} (BA18 hereafter) proposed a resolution booster for high resolution spectrographs. The FIOS (Fabry Perot Instrument for Oxygen Searches) concept is a Fabry Perot Interferometer (FPI) array that creates an ultra-high resolution spectral transmission profile. When coupled to an existing high-resolution spectrograph, FIOS can boost the spectral resolving power of the instrument to the one optimal for \mo{} detection. A lab demonstrator of FIOS with two etalons/dualons was presented in \cite{Rukdee_2020} (Rk20 from here on). In this work, we describe the first on-sky observations of the Sun using the FIOS lab demonstrator while coupled to a solar telescope. Our experimental setup and methodology are presented in section 2. Data reduction is discussed in section 3, and results are presented in section 4. We conclude this publication with a discussion in section 5 and a summary in section 6. 

%--------------------------------------------------------------------
\section{Setup \& Methodology} \label{sec:concept}

In the following section, we present a lab demonstrator of a Fabry Perot array instrument, and on-sky measurement using a solar feed. Consequently, we discuss the theoretical parameters of the instrument and present measured properties, with the goal of demonstrating the capabilities of a FIOS-like instrument to achieve extremely high resolution at the \mo{} A-band while maintaining high throughput.

\subsection{Instrument Concept and Laboratory Prototype}

\begin{figure}[h]
\includegraphics[width=0.99\columnwidth]{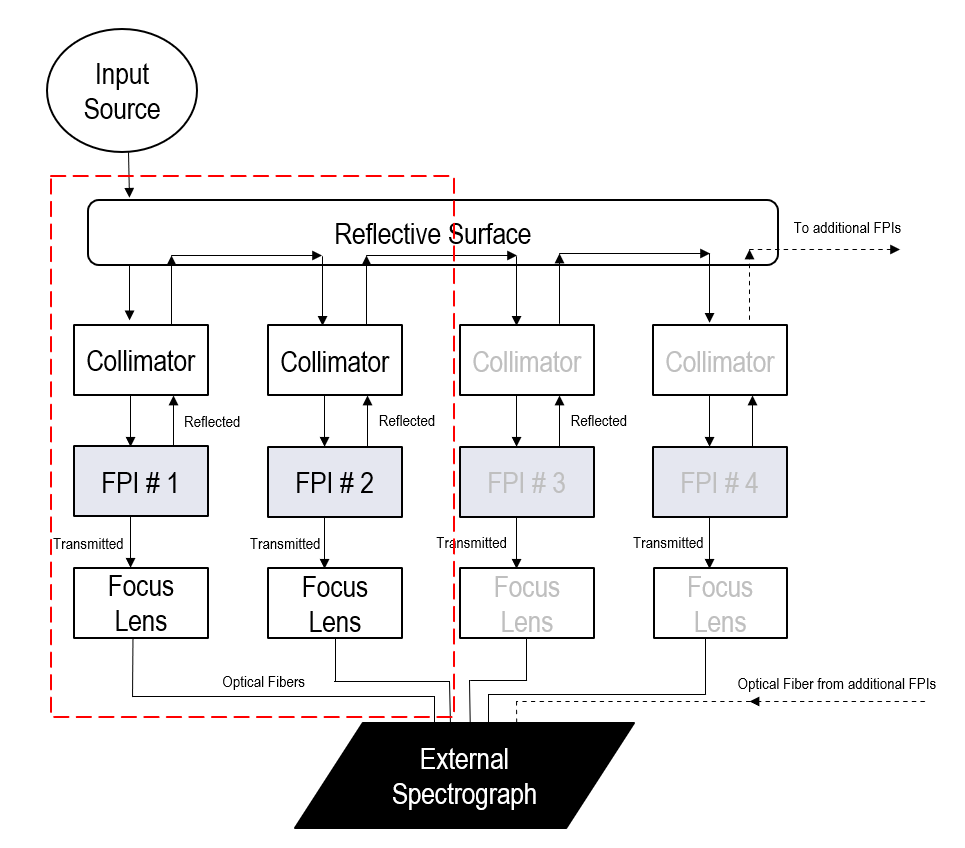}
\caption{Block Diagram of FIOS concept adapted from \cite{Ben_Ami_2018}. Light enters the system at the top left via an optical fiber. Each chain in the array consists of a collimator, an FPI and a focusing lens. Part of the light reflects from each FPI and is redirected at the central reflective surface into the next chain. The chained spectral profiles are created and fed into the external spectrograph using optical fibers. The red dashed box in the block diagram shown in this figure is equivalent to the red dashed box of the FIOS demonstrator in Figure \ref{fig:systemcartoon}.}
\label{fig:block_diagram}
\end{figure}

FIOS is a novel instrument concept, which utilizes a FPI (also called etalon or dualon, a multi-mirror etalon) to generate a transmission comb with a spectral resolution and sampling frequency well in excess of R=100,000 (BA18). The high-resolution, yet continuous transmission comb is achieved by chaining several etalons together, each having a slightly different thickness so the transmission profile of each etalon is shifted in wavelength space by the desired spectral resolution. This is repeated for N etalons in the array. N is given by the ratio of the target resolution to the free spectral range (FSR), a distance between two peaks of the etalon spectral profile. The reflected beam from the Nth etalon in the array is directed to the N+1th etalon. The beam transmitted through the Nth etalon is fed to an external spectrograph with a spectral resolution higher than the etalon's FSR in order to separate overlapping interference orders. This chain FPI array concept is illustrated in Figure \ref{fig:block_diagram}.

\begin{table}
\renewcommand{\thetable}{\arabic{table}}
\centering
\caption{Etalon \& Dualon Specification} \label{tab:Dualonspec}
\begin{tabular}{lll}
\hline
\hline
%\decimals
& Etalon & Dualon \\
\hline
Manufacturer &  \multicolumn{2}{c}{LightMachinery} \\
Absolute thickness &  26.308  mm  & 26.308 mm \\
Side Length & 25.36 mm &  25.4  mm   \\
Dualon Mismatch & n/a & 135.0 nm \\
Int. Surface Reflectivity & n/a & 93.89-94.66\%  \\
Ext. Surfaces Reflectivity & 72.5\% &  63.1\%  \\
%Calculated finesse &   9  \\
Coating &   750-780 nm &  750-780 nm \\
Edge square to mirrors &  2.9'  & $<$ 5'  \\
\hline
\end{tabular}
\end{table}

\begin{figure}[h]
\includegraphics[width=0.99\columnwidth]{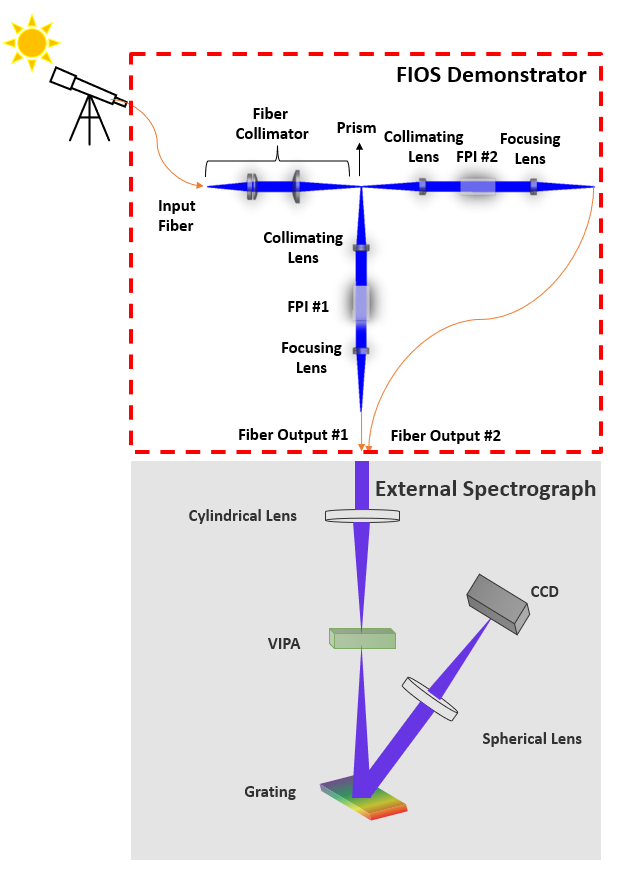}
\caption{An illustration of the light path (blue) through the two-arm FIOS prototype and accompanying external spectrograph (not to scale). On the top left, the telescope feeds the system with a fiber (orange). Top: The FPI two-arm set-up is used to hold two etalons/dualons and create the ultra-high resolution spectral profile that is fed to an external spectrograph. Bottom: LightMachinery external spectrograph.}
\label{fig:systemcartoon}
\end{figure}

The FIOS instrument concepts presented in BA18 consists of eight etalon arm chained together. As a proof of concept, Rk20 built and characterized a two-arm lab demonstrator for the etalons and dualons. Rk20 also quantified the throughput and resolution gains when replacing the etalons for dualons in the prototype. Dualons are three mirror etalons with flat-topped transmission curves and higher out-of-band rejection capabilities. Figure \ref{fig:systemcartoon} shows the prototype layout. The FPI array is shown on the top box labeled as FIOS demonstrator. The etalon and dualon specifications are listed in Table \ref{tab:Dualonspec}. The first and second etalon/dualon in the chain differs slightly in thickness (so called mismatch) so that the transmission combs are shifted by the target resolution of R=500,000. From the lab prototype with two arms, Rk20 achieved a resolving power of approx. R = 350,000 at 765 nm and a maximum throughput of  $80\%$ in the 750-780 nm  band through the etalon array. When using dualons in lieu of etalons, Rk20 achieved a $\sim10\%$ higher overall throughput and $15\%$ higher resolution using identical input beams. Motivated by this work, we employed dualons in our experimental setup.

\subsection{Setup and On-Sky Observations}

We demonstrated the capabilities of our FPI array by recording telluric features in Earth's atmosphere during daytime. Our setup consisted of three subsystems: a solar fiber feed; an FPI array (FIOS demonstrator); and a cross-disperser. Each subsystem is described below.

\begin{figure}[]
\includegraphics[width=0.99\columnwidth]{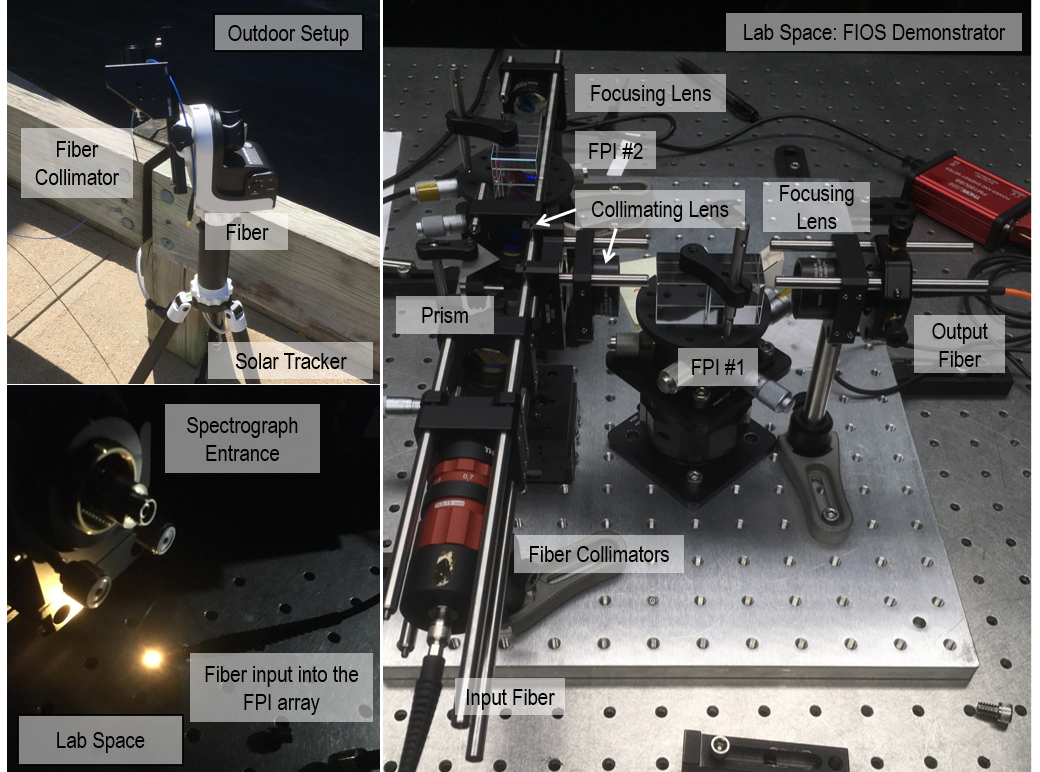}
\caption{The solar feed and optical fiber (Top left panel) channels sunlight into the lab (Lower left panel). For illustration, the fiber-end is not plugged into the FPI input port. The input fiber is then plugged into the FIOS demonstrator (Right panel) with the two-arm FPIs (dualons) }
\label{fig:solar_tracker}
\end{figure}

The solar fiber feed consisted of a narrow band filter ($\Delta\lambda\sim10\,$nm around $760$nm) and an objective mounted on a solar tracker and coupled to a $50\,\mu$m multi-mode Polymicro fiber with an NA$=0.22$ (Figure \ref{fig:solar_tracker}). We placed the solar tracker at the loading dock of our lab, with the fiber routed to the input port of our FPI array.

\begin{figure}[h]
\includegraphics[width=0.99\columnwidth]{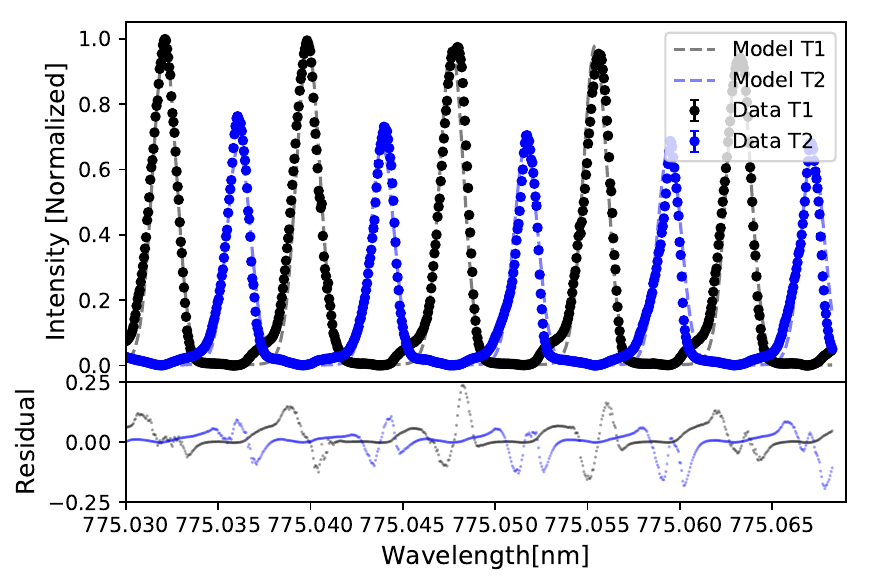}
\caption{The chained FPI profiles fed into the external spectrograph. The black dataset is obtained from the first arm and the blue one is from the second arm. Both datasets are normalized and fitted with the FPI model and Gaussian convolution. \label{fig:fpi_spectra}}
\end{figure}

Rays entered the FPI array at the fiber input. We calibrated and aligned the array prior to observations using a tunable laser and a fiber of identical properties. We used the tunable laser and a power meter to sweep wavelength space (750-780 nm) and measured the intensity of the beam transmitted by each dualon in the array using a power meter. We estimated a measurement error of $\sim1\%$ by monitoring the intensity from the input optical fiber directly for several minutes using the same power meter before each measurement. Figure \ref{fig:fpi_spectra} shows the characteristic comb-like profile of the transmitted spectra from each arm of the FPI array. Each peak's full width half maximum (FWHM) corresponds to a resolution of R = 350,000 with the 50 $\mu$m fiber input. The signal obtained from each arm was normalized to the first arm maximum. To measure the incident angles of each FPI, we fitted the data with the same FPI model as in Rk20. The best-fit incident angles are $0.103^\circ$ and $0.108^\circ$ (approximately $6\,$arcmin) for the first arm (T1) and second arm (T2) respectively. This is slightly larger than the $5\,$arcmin divergence of the input beam following collimation, which we measured with a beam profiler.

\begin{figure}[h]
\includegraphics[scale=0.4]{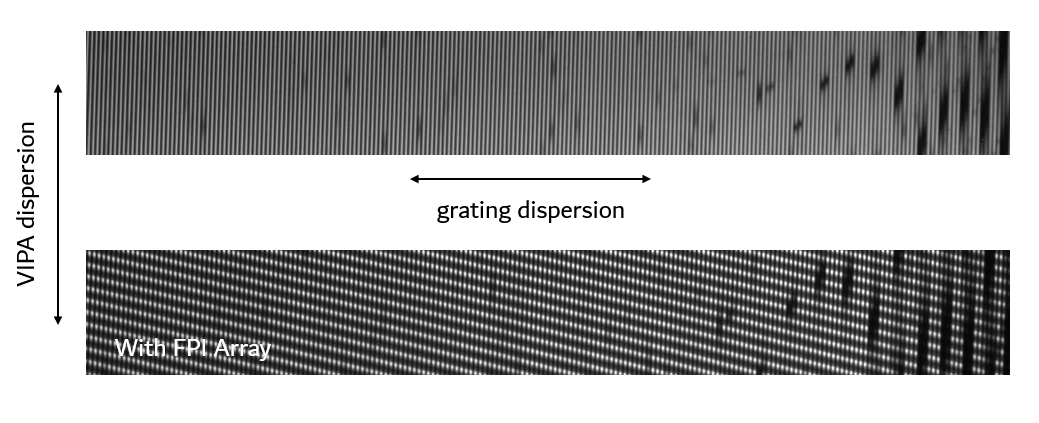}
\caption{Solar spectra obtained by the VIPA external spectrograph. Top: detector image without FPI profiles showing dispersion in the vertical direction; Bottom: detector image with imposed FPI profile from top left to the bottom right direction with solar absorption. \label{fig:ccdimages}}
\end{figure}

The last component in our setup is an external spectrograph. Here we use the HyperFine Serie from LightMachinery Ltd., as a cross-disperser for the chained FPIs. The spectrograph employs a Virtually Imaged Phased Array \citep[VIPA;][]{Shirasaki1996} dispersive element designed to measure hyperfine spectra and small spectral shifts. The VIPA functions similarly to a ruled grating and produces extremely high dispersion along a predefined axis. Overlapping orders are dispersed in the perpendicular axis using a ruled grating positioned downstream. The spectrograph was tuned to operate in the 760-780 nm wavelength range. In our setup, the transmitted beam from each FPI in the array was coupled to the external spectrograph individually. A fully operational FPI array instrument should be coupled to an external spectrograph that can accept several fibers simultaneously (\textit{e.g.}, G-CLEF  \citealt{2016SPIE.9908E..22S}). We set the integration time for each exposure to be 12 seconds when fed with our solar feed, determined empirically. This allowed us to avoid detector saturation while maintaining high SNR in each exposure.  The result was a 2D spectrum of the input beam as shown in Figure \ref{fig:ccdimages}. 

The top panel of Figure \ref{fig:ccdimages} shows the dispersed image from the external spectrograph alone (without the FPI unit) of a solar spectrum. This partial detector image is similar to those obtained from a high-resolution echelle spectrograph in the horizontal axes. Absorption lines of telluric oxygen are clearly visible as vertically elongated dark spots. The bottom panel of Figure \ref{fig:ccdimages} shows the dispersed image from the external spectrograph and the FPI unit. The repeating dualon profile (Figure \ref{fig:fpi_spectra}) is imposed in Figure \ref{fig:ccdimages}, in the bottom left to top right direction. As can be seen, the signal is modulated by the transmission profile of the FPI in the array. While we intend the FPI array to be a resolution booster, in our setup the VIPA spectrograph offers a high enough resolution to resolve the FPI transmission profile, with a spectral resolution of $3-4\times $ the dualon's FSR. We mimic the behavior of an external spectrograph whose resolution is high enough to separate transmission peaks for a given dualon by integrating over a dualon FSR. In a final configuration, an external spectrograph of lower resolution should be used to take advantage of the FPIs in the array as the elements dictating the spectral resolution of the instrument.

\section{Data Reduction}
In this section, we extracted the spectrum and analyze the \mo{} A-band feature at ultra-high resolution from five datasets. The data obtained from our prototype had a strong instrumental profile imposed by the transmission properties of each dualon, see Figure \ref{fig:fpi_spectra}. Calibrating out this profile is a crucial step before analyzing the spectrum.  

\begin{figure*}
\begin{center}
\includegraphics[width=1.0\textwidth]{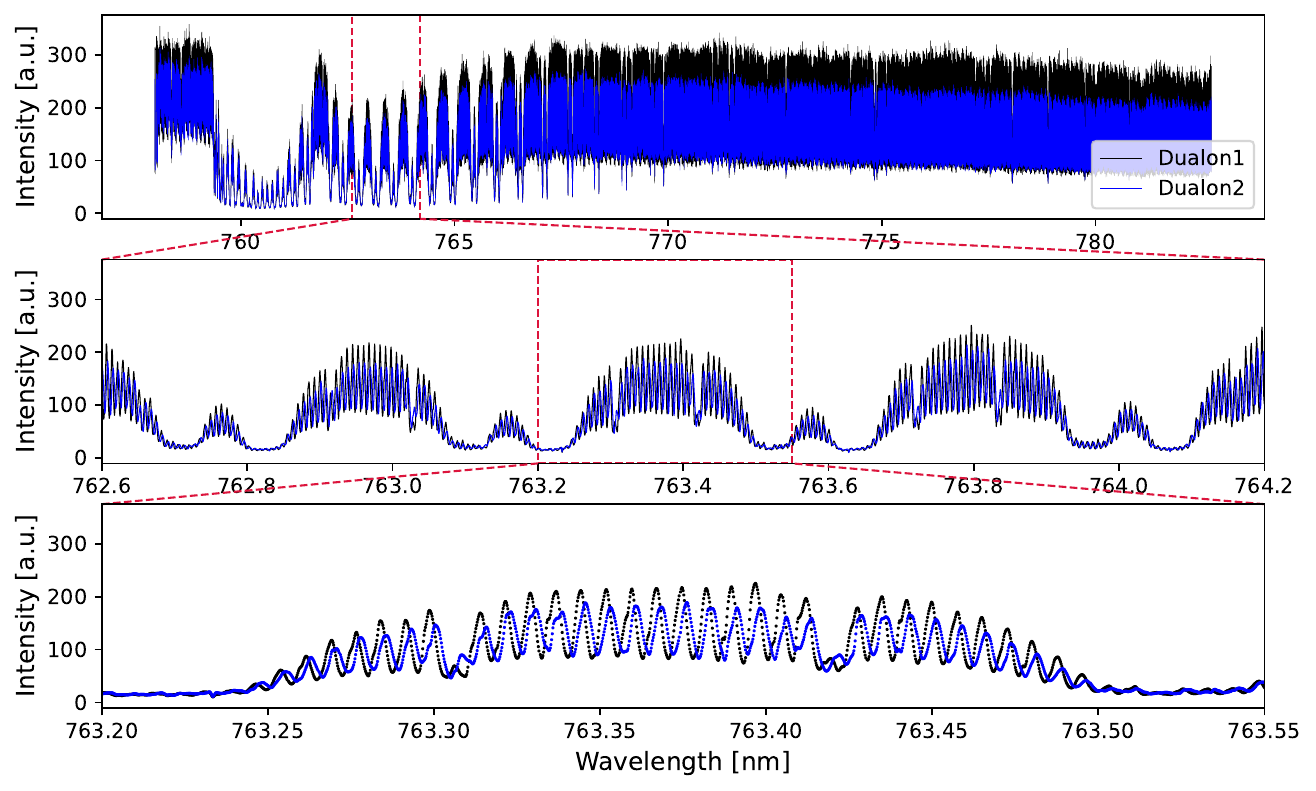} 
\caption{Top: the oxygen A-band observation using two dualons in the arrays: dualon1 data are displayed in black solid line and dualon2 in blue. Middle: the zoomed-in region shows the features of the chained spectrum both the absorption feature embedded on top of the instrumental profiles. The signal of the dualon2 is scaled up by a factor of 1.5. Bottoms: the zoomed-in region from the middle panel shows the shift between the two dualons.}
\label{fig:zoom2arms}
\end{center}
\end{figure*}

Raw images captured by the HyperFine external spectrograph were reduced using SpectraLok, a proprietary software from LightMachinery. Flat field and wavelength solutions were obtained by performing an initial calibration of the spectrograph with a quasi-uniform broadband source (namely, a laser-driven light source and a Halogen lamp), and an emission line source (Krypton arc lamps). The reduction pipeline also performed background subtraction and baseline correction. The pipeline included an additional step, the stitching together of successive spectral orders of the VIPA dispersion profile. The top and middle panels in Figure \ref{fig:zoom2arms}  show the spectrum extracted from the external spectrograph, with no correction for the dualon profiles. The black and blue curves show the spectra obtained through the first and second dualons in the chain followed by dispersion from the external spectrograph. The bottom panel shows the shift between the two arms due to a mismatch between the two dualons designed to be a few nanometers apart. The data exhibited the quasi-equally-spaced comb-like structure, which corresponded to the ultra-fine transmission structure R=350,000. The instrumental profiles can be described using a simple model for the transmission and reflection intensity profiles from \cite[e.g.,][]{Vaughan}:

\begin{equation}
I_T = \frac{T^2}{(1 - R)^2(1 + F \sin^2 (\frac{\phi}{2}))}
\label{eq:I_T}
\end{equation}

\begin{equation}
I_R = \frac{F \sin^2 (\frac{\phi}{2})}{1 + F \sin^2 (\frac{\phi}{2})}
\label{eq:I_R}
\end{equation}

\noindent
where T and R are the surface intensity reflection and transmission coefficients, respectively. The phase lag is the phase difference caused by the optical delay for successive reflections:
\begin{equation}
\phi = \frac{2\pi}{\lambda}\times 2 d\times n_\lambda\times\cos\theta
\label{eq:phi}
\end{equation}
Here $d$ is the separation between the two surfaces of the etalon, $n$ is the refractive index depending on the wavelength and $\theta$ is the angle of incidence. 
Following BA18, an additional degradation of $\theta$ due to imperfect collimation up to $\delta\theta_{max}$, can be taken into account assuming $\delta\theta_{max} = \pm\Phi_{fiber}/2f_{col}$, where $\Phi_{fiber}$ is the fiber diameter, $f_{col}$ is the focal length of the collimating lens. The phase lag term $\delta\phi$ is approximated by:

\begin{equation}
\delta\phi_\pm = \frac{4\pi}{\lambda}dn_\lambda\ (\sin(\theta\pm\frac{\Phi_{fiber}}{2f_{col}})-\sin\theta)
\label{eq:beam deviation}
\end{equation}
To describe the phase lag, we conservatively adopt $\phi'=\phi+|\phi_+|+|\phi_-|$ in place of $\phi$.

\begin{figure*}[ht!]
\begin{center}
\includegraphics[width=1.0\textwidth]
{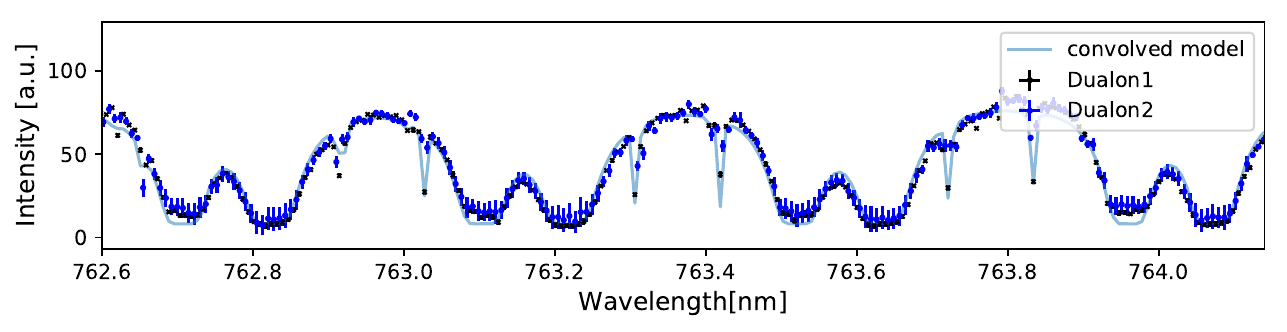} 
\caption{The reduced data with the dualon profile removed. The black data points are from dualon 1 and the blue ones are from the dualon 2. The blue error bars are larger than the black ones due to the relative intensity observed from each arm. The convolved model of the oxygen A band is shown in light blue solid line.}
\label{fig:reduced_data}
\end{center}
\end{figure*}

The finesse $F$ in equations~\ref{eq:I_T} and \ref{eq:I_R} describes the resolution boost. It is the ratio of the FSR to the width of a single transmission peak. As explained in section 2, the reflected light from the first arm $I_{R_{1}}$ was redirected into the second arm and was thus the input signal for the transmitted and reflected light of the second arm. In the full FPI chain, the transmitted signal through the $n$th arm can be described recursively:

\begin{equation}
I_{T_n} = I_{R_{n-1}} \frac {T_{n-1}^2}{(1-R_{(n-1)})^2(1 + F \sin^2 (\frac{\phi}{2}))} 
\label{eq:I_T2}
\end{equation}
The recursion starts from $n=2$ with coefficients $T_1=T$, $R_1=R$ and reflection intensity $I_{R_1}=I_R$ from equations~\ref{eq:I_T} and \ref{eq:I_R}.

In this work, we measure $I_T$ from dualon 1 and $I_{T_2}$ from dualon 2.

The next step in the data reduction is determining the locations of the dualon transmission peaks and transmission profiles. 
These are parameterized by reflectivity $R$, reflective index $n_\lambda$, and for each FPI, thickness $d$, incident angle $\theta$, and the skew-gaussian convolution standard deviation $\sigma$, skew $\alpha$ and normalization $N$.

The parameters are determined by a $\chi^2$ fit of the model to data in the $761-770\,$nm bandpass where the central wavelength of the \mo{} ($\sim765\,$nm) is located. We limit the fit to the continuum regions to avoid absorption lines. A further model modification is necessary, as the FSR increases monotonically, but non-linearly, with wavelength. Without accounting for this, the best-fit periodic model runs out of phase with the periodic observed signal after a few hundred cycles. Following \cite{Zhu_2020}, we adopt a 4th order polynomial wavelength solution. For stabilizing the fit, we first used a narrow wavelength range (0.5-1 nm) window starting at  761.0 nm, with the higher order polynomial coefficients initialized at zero. Then, we iteratively extend the wavelength fitting window. The four polynomial coefficients, and the five dualon parameters for each arm, are well determined by the $\sim50,000$ data points used in the fit, and so we obtain a model of the two dualon arms across the entire observed bandpass. 

Finally, the reduced spectrum is obtained by marginalizing out the FPI profile. Figure \ref{fig:reduced_data} shows the spectrum with the dualon profiles calibrated out as a comparison to the middle panel of Figure~\ref{fig:zoom2arms}. This is achieved by integrating the flux within each FPI profile model transmission peak. This produces one data point per FSR. It does not result in loss of resolution, as the wavelength range is confined to a significantly narrower bandpass within the data point. We simulate the behavior of an external spectrograph with a resolution similar to that of the dualon FSR, as described in BA18. The integration over the peak is weighted by its transmission profile (Figure~\ref{fig:fpi_spectra}). The procedure is applied to each $12\,$sec exposure independently. Then we take an average of five datasets of $12\,$sec exposure each and obtain the fitted model's mean and standard deviation for each data point. Between observations, the dualon transmission profiles shift by a small fraction of the transmission peak FWHM width, in average $0.2\,$pm, an order of magnitude below the FWHM of a transmission peak, which permits the co-addition of the extracted spectra. 

\section{Results} \label{sec:results}

Once the instrumental profiles were marginalized, we calibrated the two datasets using a spectral model of telluric oxygen. Then the FWHM of the spectral feature was determined as described in this section. 

\subsection{Spectral resolution}
After marginalizing out the instrumental profile, we estimated the spectral resolution of the lines in the \mo{} band in two steps: 1) calibrating the two dualon profiles with $\chi^2$ fit of a telluric model, and 2) measuring the FWHM of detected lines. Our FWHM was derived from the standard deviation of the fit. In the first steps, the two datasets were fitted to a model of telluric absorption at various spectral resolutions. For the narrow wavelength window considered, a wavelength-independent intrinsic solar spectrum  $F_\mathrm{in} = A$, was assumed and propagated through a model of the telluric lines, $F_\mathrm{out}$, given by:

\begin{equation}
F_\mathrm{out}  = A \times \underbrace{\exp\left(-\tau * \sigma_{\lambda}\right)}_{T(\lambda)}
\end{equation}

\begin{figure*}[]
\includegraphics[width=\textwidth]{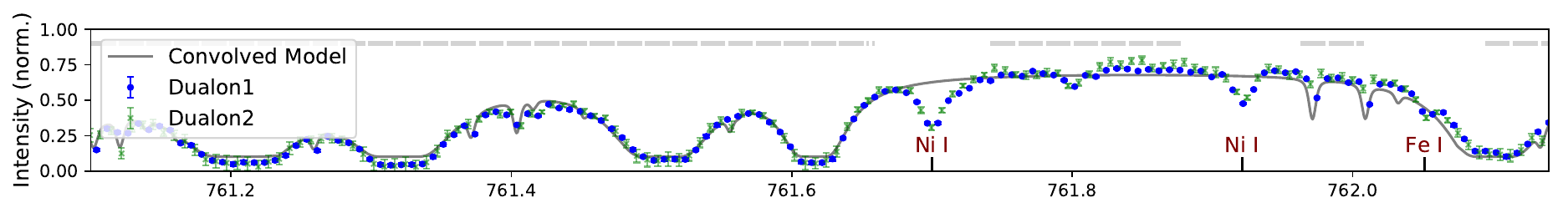} 
\includegraphics[width=\textwidth]{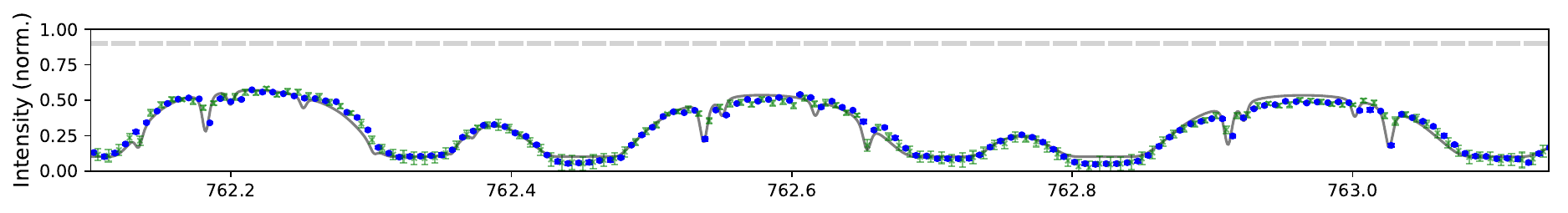} 
\includegraphics[width=\textwidth]{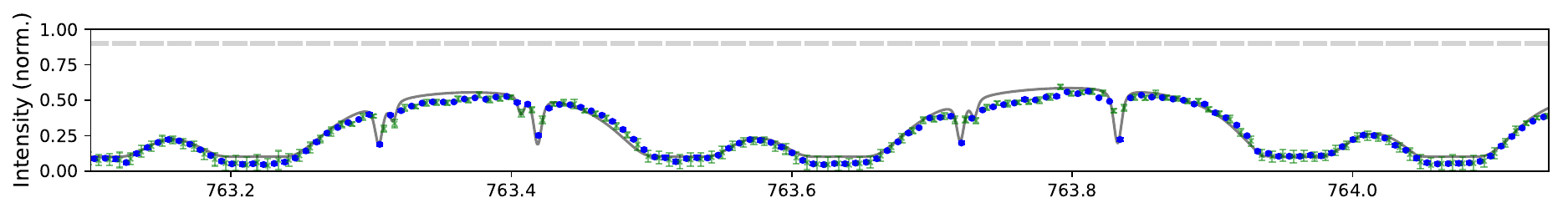} 
\includegraphics[width=\textwidth]{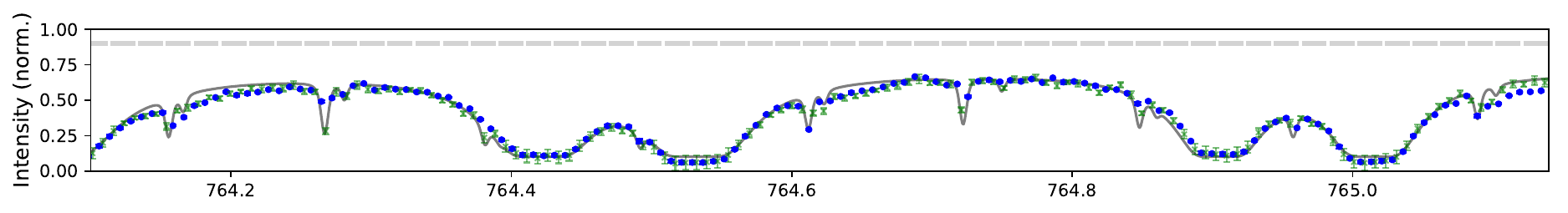} 
\includegraphics[width=\textwidth]{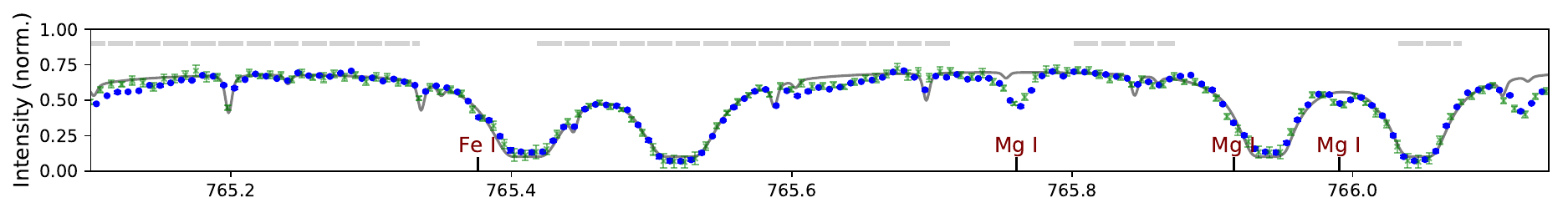} 
\includegraphics[width=\textwidth]{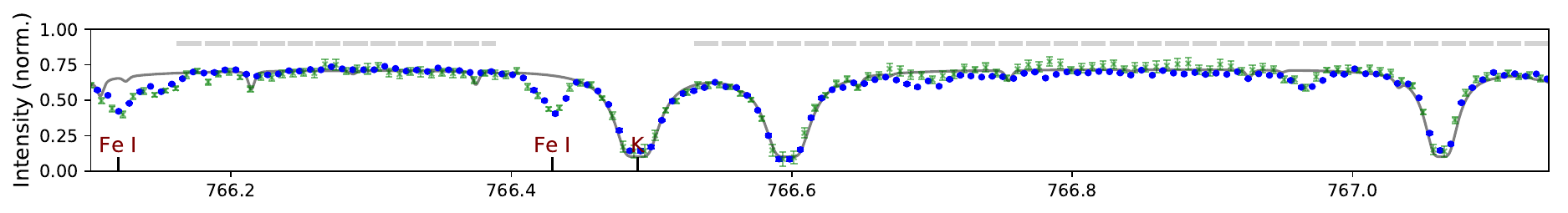}
\includegraphics[width=\textwidth]{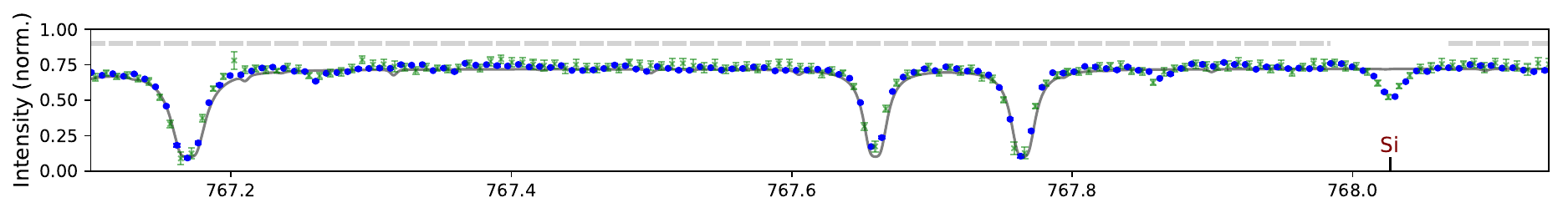} 
\includegraphics[width=\textwidth]{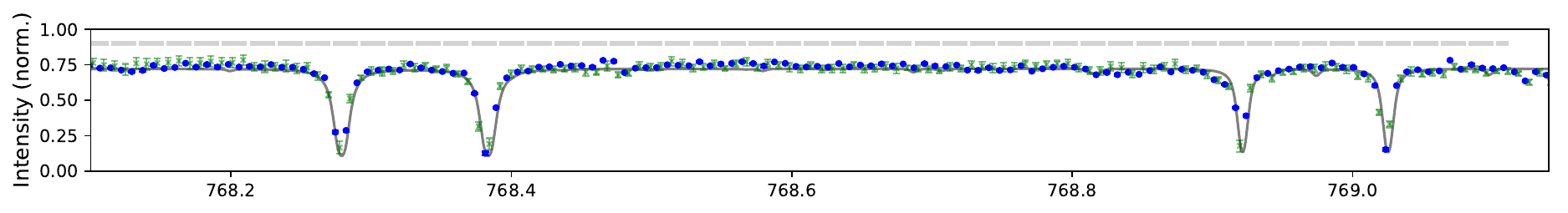} 
\includegraphics[width=\textwidth]{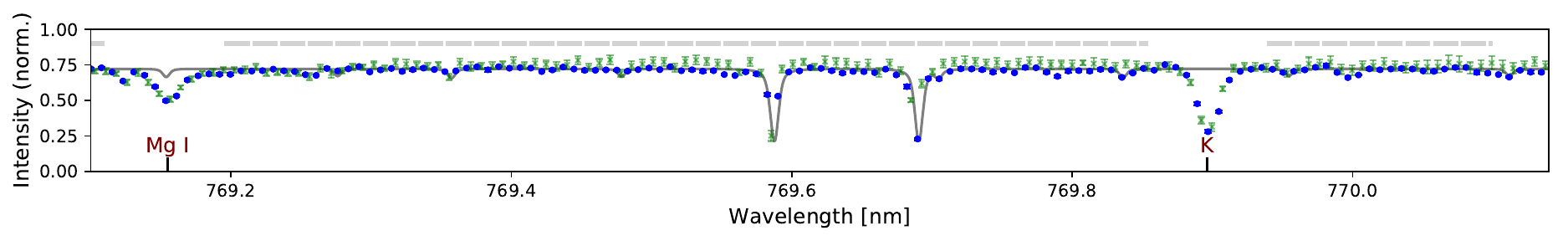} 
\caption{The reduced data points from the first and second arm of the FPI unit (indicated as dualon 1 (blue) and dualon 2 (green)) fitted with model of the oxygen A-band, where the central wavelength lies within the window $760-770\, nm$. The non-fitted features are marked as the atomic absorption lines from the solar spectrum. \label{fig:elements}}
\end{figure*} 

The telluric absorption model $T(\lambda)$, was taken from ESO SkyCalc \citep{Noll_2012,Jones_2013,Moehler_2014} and describes the attenuation cross-section $\sigma_{\lambda}$, at a nominal optical depth, $\tau$. We allowed the optical depth to vary by raising $T$ to the power of absorption depth, $n$. Small wavelength shifts $\Delta\lambda$ were allowed to match the wavelength solution of the data and the model; they differed by 0.26 nm on average according to the fitted value. This shift matched the wavelength of our observed spectra to the model template. Lastly, a constant background contribution (background noise), $B$, was added. With these modifications, we defined the system output flux as follows: 

\begin{equation}
F_\mathrm{out} = A \times T(\lambda+\Delta\lambda)^n + B
\end{equation}

\noindent
The native resolution of the telluric absorption model $T(\lambda)$ was R=1,000,000. We derived telluric absorption models at various resolutions by convolving $F_\mathrm{out}$ with a set of Gaussian kernels of various width, $\sigma$, to degrade the spectral resolution:
 
\begin{equation}
F_\mathrm{model}  =  (A \times T(\lambda+\Delta\lambda)^n + B) * \mathrm{N}(0,\sigma)
\end{equation}

\begin{figure}[ht!]
\begin{center}
\includegraphics[width=1.0\columnwidth]{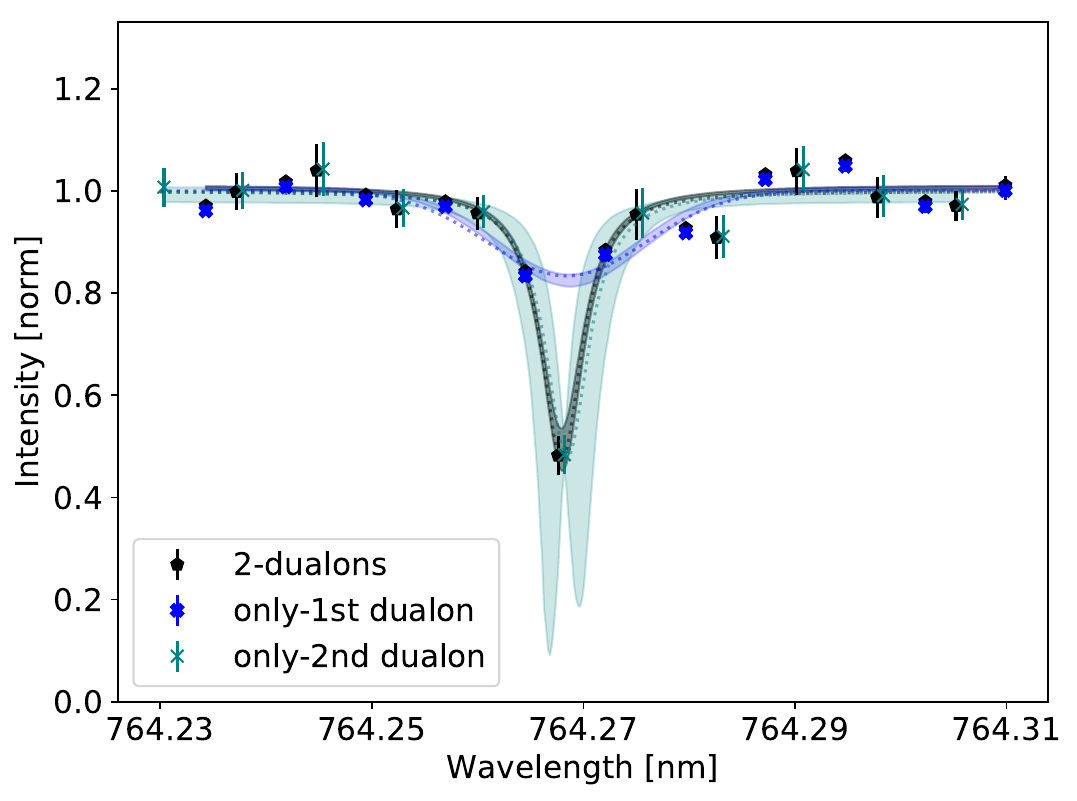}
\includegraphics[width=1.0\columnwidth]{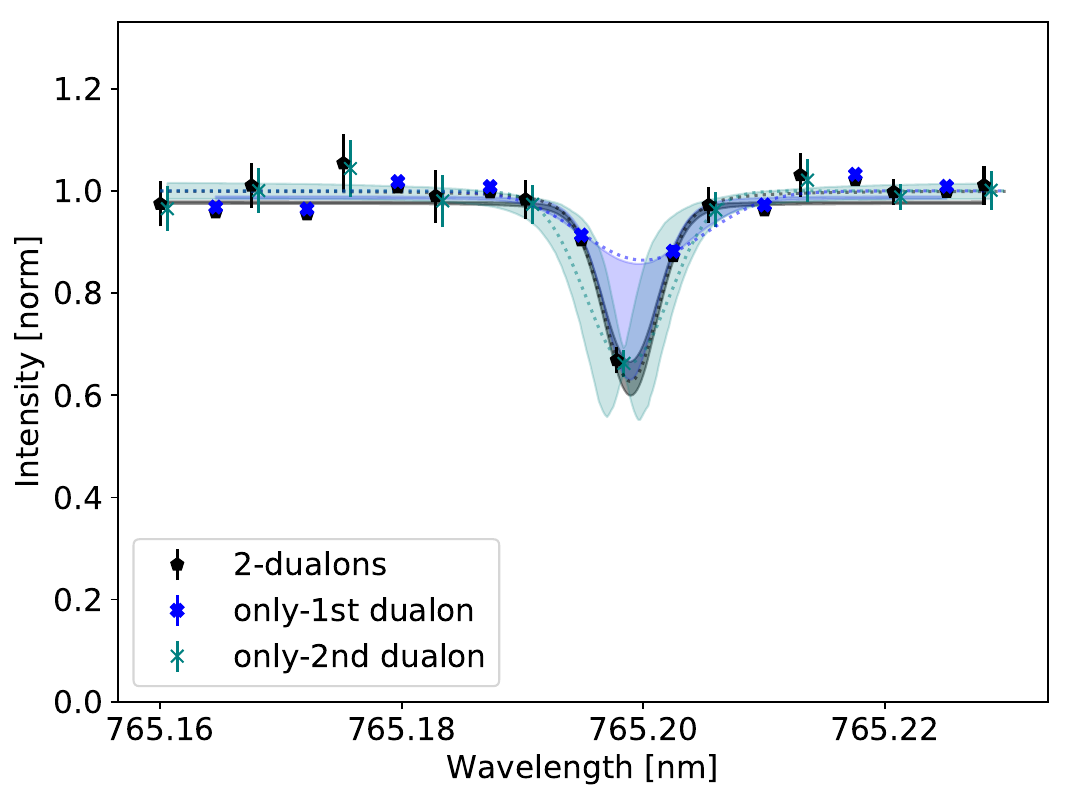}
\includegraphics[width=1.0\columnwidth]{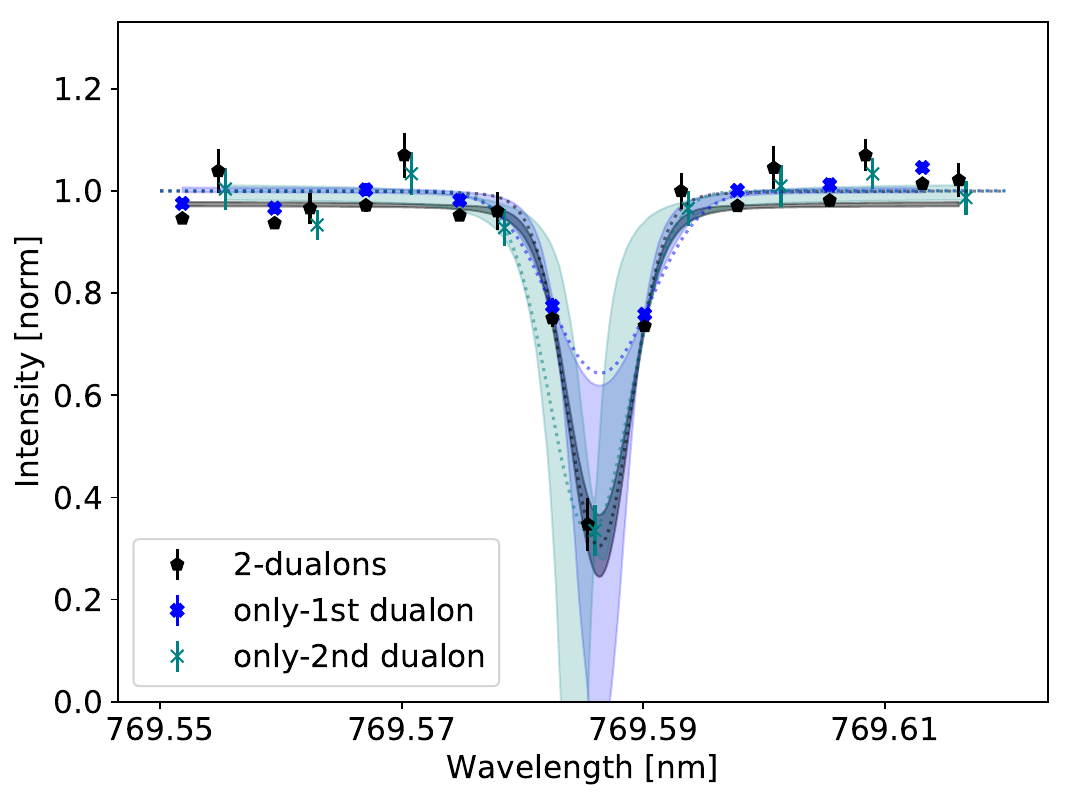}
\end{center}
\caption{The zoomed regions within the \mo A-band of three datasets: 2 dualon (black), 1st dualon (blue) and 2nd dualon (green). Each dotted line is isolated and fitted with a Voigt model. The shaded area is the uncertainty of the fit. The black, blue and green transparent shading correspond to the uncertainty of fit for the 2-dualons, only-1st dualon and only-2nd dualon respectively. The FWHM results are reported in Table \ref{tab:R_compare_only_dualons}.  \label{fig:R_compare_only_dualons}}
\end{figure} 

\noindent
The Gaussian standard deviation $\sigma$, absorption depth $n$, amplitude $A$, background $B$ and wavelength shift $\Delta \lambda$ were parameters of the $\chi^2$ fit and then convolved with a Gaussian with standard deviation, $\mathrm{N}(0,\sigma)$. The data from the second arm were fitted jointly with the first arm, appropriately taking into account transmission losses between the two arms. The full high-resolution fitted spectrum over the $760-770\,nm$ waveband is shown in Figure \ref{fig:elements}. The horizontal gray dashed line at the top of each panel indicates the fitted region to the model. The atomic line regions of the solar abundances were masked out from this fit. These features are described below. The data points from the second arm (circles) followed the first arm (asterisks) alternately. To determine the respective background and efficiency, each arm had independent free parameters for A and B. The two arms shared power and standard deviation, while the amplitude, constant background from each dataset and wavelength shifts are different. This allowed us to normalize the data from the two arms relative to each other for subsequent analysis. However, we did not use the FWHM (or convolution $\sigma$) from this analysis, because the model broadening is too simplistic for high-resolution analyses. With the normalized joint data set from both arms, we analyzed individual narrow absorption features to evaluate the spectrograph performance. The data reduction process and the fitting procedure discussed above were required to normalize the two arms and determine the continuum model.

Once the spectral features were located, we measured the FWHM of a few spectral lines. For the analysis of the line-width, we aimed to determine the resolution by measuring the FWHM using single narrow telluric absorption lines. The SkyCalc model has a R=1,000,000. Employing the line-width approach enabled the comparison of resolution performance between dualons, and comparison with other instruments when employing two dualons together. A Voigt model with parameters: amplitude, mean, standard deviation and background, was fitted with MCMC using the python \textit{emcee} \citep{Foreman_Mackey2013} open source package, allowing us to determine the uncertainty in each wavelength window. Then, the corresponding FWHM is reported in Table \ref{tab:R_compare_only_dualons}. Both dualons are needed to fully map the valley of the \mo{} lines. Together they improve the overall resolution by a factor of two as reported in Table \ref{tab:R_compare_only_dualons}. Figure \ref{fig:R_compare_only_dualons} shows the results from this analysis. The smallest telluric feature observed through this setup (50 $\mu$m fiber + dualon + Ext. spectrograph) is resolved at $\Delta\lambda$ = 4.5 pm.

\begin{table}[]
\renewcommand{\thetable}{\arabic{table}}
\centering
\caption{Selected spectral region for the resolution comparison of the dualons. All units are in nm.}  \label{tab:R_compare_only_dualons}
\begin{tabular}{cccc}

\hline
& FWHM & FWHM & FWHM \\
$\lambda_{peak}$ & 2 dualons  &  1st dualon & 2nd dualon  \\
%[nm] & - & -& [nm] & - & [nm]  \\

\hline
\hline

764.27 & 0.0046 $\pm$ 0.0004 & 0.0123 $\pm$ 0.0012 & 0.0048 $\pm$ 0.0020 \\
765.19 & 0.0054 $\pm$ 0.0002  & 0.0149 $\pm$ 0.0002 & 0.0063 $\pm$ 0.0024 \\ 
769.58 & 0.0061 $\pm$ 0.0004 & 0.0071 $\pm$ 0.0024 & 0.0091 $\pm$ 0.0002\\

\hline
\end{tabular}
\end{table}

Besides determining the instrument resolution within the \mo{} lines, a few measurements were carried out with the calibration source (see section 2). The FWHM ($\Delta\lambda$) measurements of the spectral lines from various sources and fiber sizes from this work are summarized in Table \ref{tab:R_budget}. The measurements were carried out with different setups and different components added into the system with different fiber sizes. First, we used a single-mode fiber with a laser source (results in Rk20) and a Krypton lamp to measure the FWHM of the spectral profile of different configurations: 1) only dualon, 2) only external spectrograph, and 3) the dualon profile plugging in the external spectrograph. The FWHM was measured to be 1.0 pm. These results agreed well among all configurations. Then we switched to a 50 $\mu$m fiber to measure only the dualon profile with a laser scan. This increased the FWHM of FIOS alone to 1.7 pm (Figure \ref{fig:fpi_spectra}). Considering the dualon together with the external spectrograph yielded a FWHM of 3.0 pm, based on a Kr lamp measurement. Comparing the single-mode fiber and the 50 $\mu$m measurement (second and fifth row of Table 3), the fiber size also degraded the performance of the external spectrograph in this case due to imperfect collimation. Finally, the narrowest telluric feature seen in our dataset were measured at 4.5 pm FWHM. The broadening from 3.0pm to 4.5 pm can be explained by the known intrinsic width of these lines e.g. the solar atlas \citep{Wallace_2011}.

\begin{table}[h!]
\renewcommand{\thetable}{\arabic{table}}
\centering
\caption{Resolution budget from different setup configuration}  \label{tab:R_budget}
\begin{tabular}{cccc}

\hline
Fiber size & Source  &  Components & $\Delta\lambda$ [pm]  \\ 
\hline
SMF & Laser & Dualon & 1.0 \\
 %& Rk20 & 760000
& Kr-lamp & Ext. Spec & 1.0 \\
 % This work  %& 760000
& Kr-lamp & Dualon + Ext. Spec & 1.0\\
 %& This work  % & 760000
50 $\mu$m  & Laser & Dualon & 1.7 \\
 %& This work  %& 450000
 & Kr-lamp & Dualon + Ext. Spec & 3.0 \\
 %& This work  %& 250000
 & Solar & Dualon + Ext. Spec & 4.5 \\
 %& This work  %& 170000
\hline
\end{tabular}
\end{table}

\subsection{Transmission}
To detect \mo{} in an Earth analog by transmission spectroscopy requires high SNR, therefore a high system transmission is crucial. Below we discuss the transmission of FIOS plugged into an external Echelle spectrograph. We choose not to address expected losses from external sources such as the atmosphere and telescope, as these are not inherent to the proposed system. 

The effective transmission of the first dualon arm as measured in our analysis is 84\% averaged across the FPI FWHM, with an increasing loss of 9\% to any additional arm due to reflection losses and an imperfect alignment in the optical path. The increased losses in the second arm are driven partly by the limited capability of the setup to direct the incident beam on the second arm with high enough precision. Zemax CAD simulations suggest that a more stable setup, with higher angular precision and stabilized environment can potentially recover 10\% transmission of the second arm. While each arm contributes data points to the final combined spectrum, the throughput progressively lowers with each arm, resulting in larger measurement uncertainties. If we take the average across the eight arms, the throughput is $\sim50$\%.

The current setup in Figure \ref{fig:solar_tracker} requires careful attention on the beam collimation, incident angle, and vignetting. All of these can lead to throughput losses. Achieving maximum system throughput requires thorough calibration and precise alignment. As the number of arms increases, the alignment process becomes more intricate, but this challenge can be addressed with high-precision kinematic mounts incorporated with dedicated stable optomechanical supports, environmental control, and the expertise gained from developing the prototype here.

Furthermore, the fiber optics system from the telescope to the FIOS unit is expected to give an additional loss of $\sim10$\%, and similar losses from the fiber optics system channeling light from the FIOS unit to the external spectrograph which allows us to separate dualon transmission orders (e.g., Espresso \citep{Pepe2014}, Maroon-X \citep{Seifahrt2022}, Express \citep{Jurgenson2016}). This sums up to a transmission of $\sim40\%$. We discuss losses due to the external spectrograph in section \ref{sec:discussion}.

\subsection{Performance Comparison to other instruments}
\begin{figure}[h!]
\begin{center}
\includegraphics[width=0.99\columnwidth]{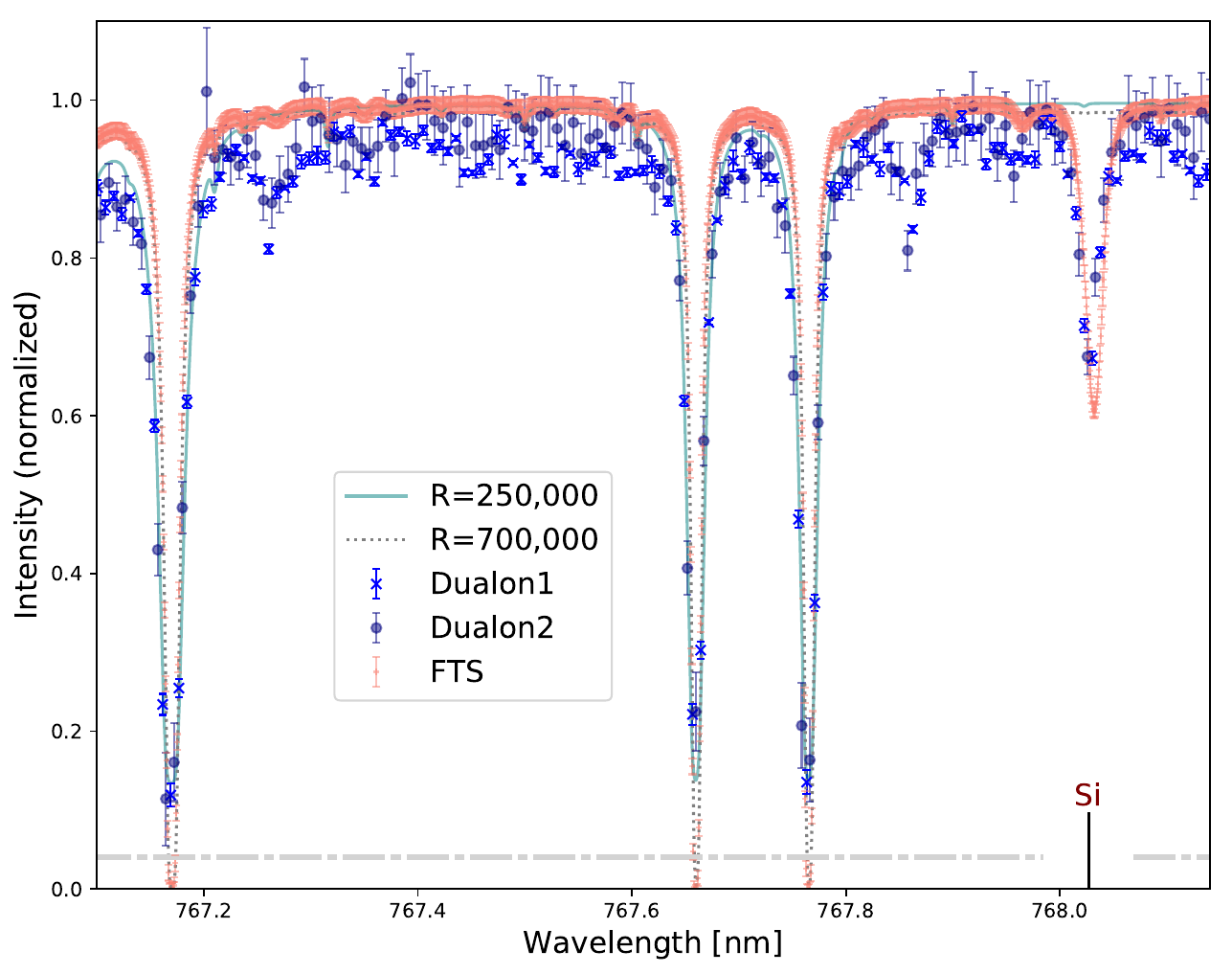}
\caption{A comparison of a zoomed region at 767 nm detected with multiple instruments: 2-arm FPI array (blue) of R=250,000, Fourier Transform Spectrograph (FTS) at Kitt Peak (red) with R = 700,000. A fitted model for the FTS data is represented by the grey dotted line, while the light blue solid line is fitted to the two dualon data. \label{fig:zoom_compare}}
\end{center}
\end{figure}

We compare three data sets: (1) our data from 2 dualons of the
FPI array with resolution ($R_{dualon}$), (2) Fourier Transform Spectrograph (FTS) from the National Solar Observatory at Kitt Peak \citep{Wallace_2011} reported $R_{FTS}$ =700,000 within our wavelength range of interest, and (3) the SkyCalc model convolved to resolution ($R_{model}$). Each line has the observed FWHM, $W_{obs}\leq0.008$ nm (R $\geq$ 100,000) from the dualon signal reported in Table \ref{tab:o2_narrow_lines_instruments}. Figure \ref{fig:zoom_compare} shows a zoomed region that compares the normalized signal from 2 dualons and FTS. The small variations observed in the dualon data within the continuum region indicate the instrumental profile of the external VIPA based spectrograph. 

To determine the instrument resolution, we take the line intrinsic width into account incorporated with the Skycalc model. Based on Bienaymé's identity, the intrinsic line width can be subtracted in quadrature from the total observed line width to obtain the instrumental width: Equation \ref{eq:sigma_obs}.

\begin{equation}
W^2_{obs}  = W^2_{int} + W^2_{LSF}
\label{eq:sigma_obs}
\end{equation}

\noindent
where $\mathrm{W_{obs}}$ is the observed FWHM, $\mathrm{W_{LSF}}$ is the FWHM of the Line Spread Function, and $\mathrm{W_{int}}$ is the FWHM of an intrinsic line. In this work, we assumed $\mathrm{W_{int}}$ from the observed FTS width. The first row of Table \ref{tab:R_compare_only_dualons} reports $\mathrm{W_{obs} = 4.6}$ pm for the 2-dualon, the FTS generally resolved at the level $\mathrm{W_{int}}=\mathrm{W_{obs}}= 3.5$ pm. Following Equation \ref{eq:sigma_obs},  therefore our $\mathrm{W_{LSF} = 2.98}$ pm corresponds to R=255,212. The other lines give similar but slightly larger R values. This number agrees with the Kr line measurement in Table \ref{tab:R_budget}.

\begin{table}[]
\renewcommand{\thetable}{\arabic{table}}
\centering
\caption{Comparison of the resolving power of some Oxygen narrow lines observed within $760-770\,\mathrm{nm}$ window observed with different instruments.}  \label{tab:o2_narrow_lines_instruments}
\begin{tabular}{cccc}
\hline
 $\lambda_{peak}$ & $R_{dualon}$  & $R_{FTS}$  &$R_{model}$ \\
\hline
763.03 & 151344  & 204053  & 184359 \\

763.83 & 155642   & 214641 & 199224 \\

764.27 & 165900  & 242737  & 222736 \\

768.91 & 100551  & 162080  & 133296 \\

769.58  & 122482  & 210896 & 181903  \\

\hline
\end{tabular}
\end{table}

\begin{figure}[]
\begin{center}
\includegraphics[width=0.32\columnwidth]{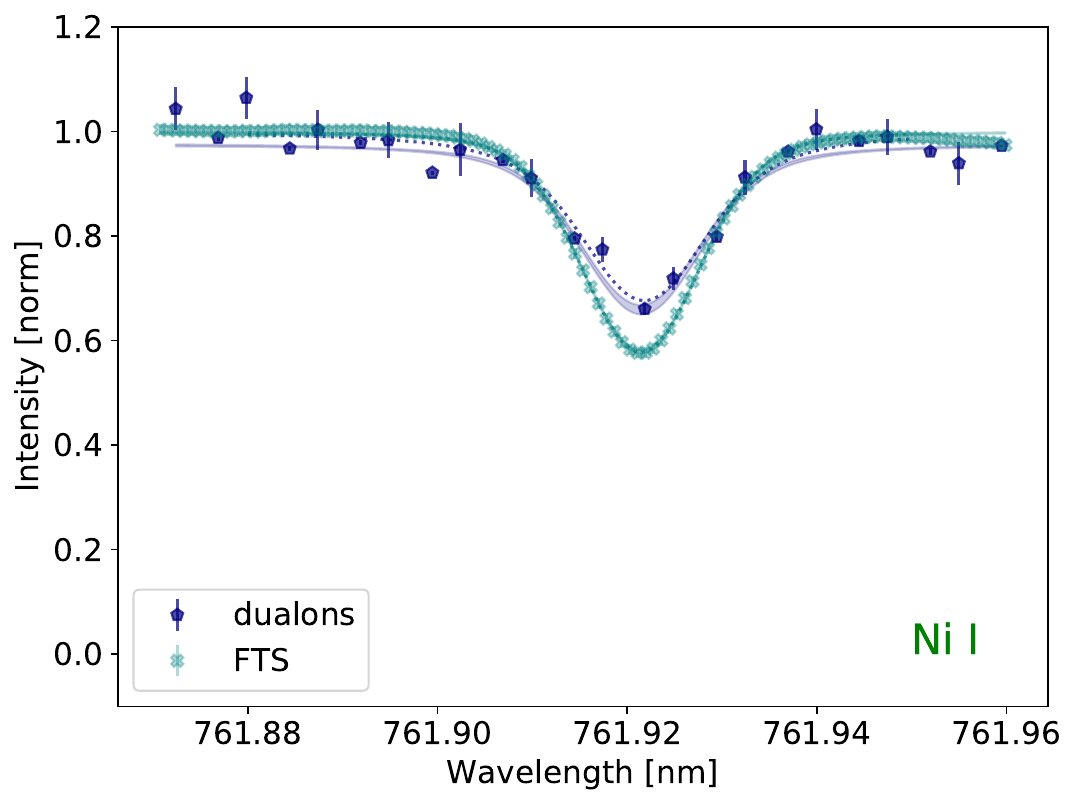} 
\includegraphics[width=0.32\columnwidth]{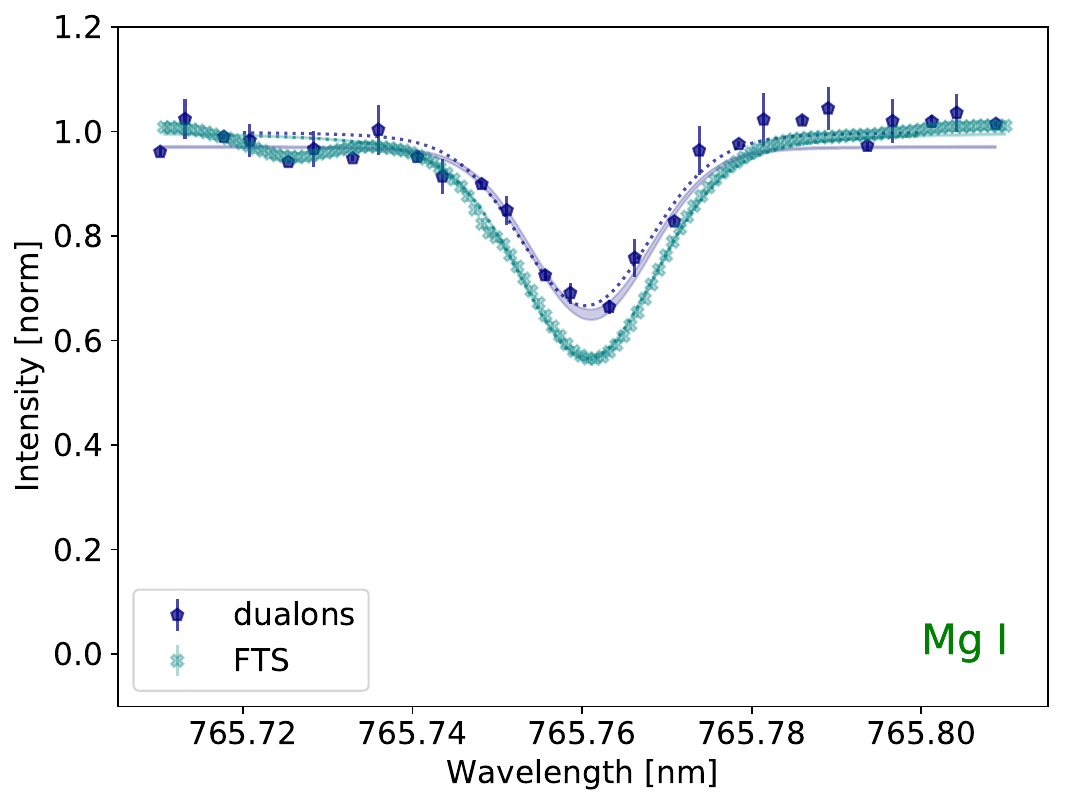} 
\includegraphics[width=0.32\columnwidth]{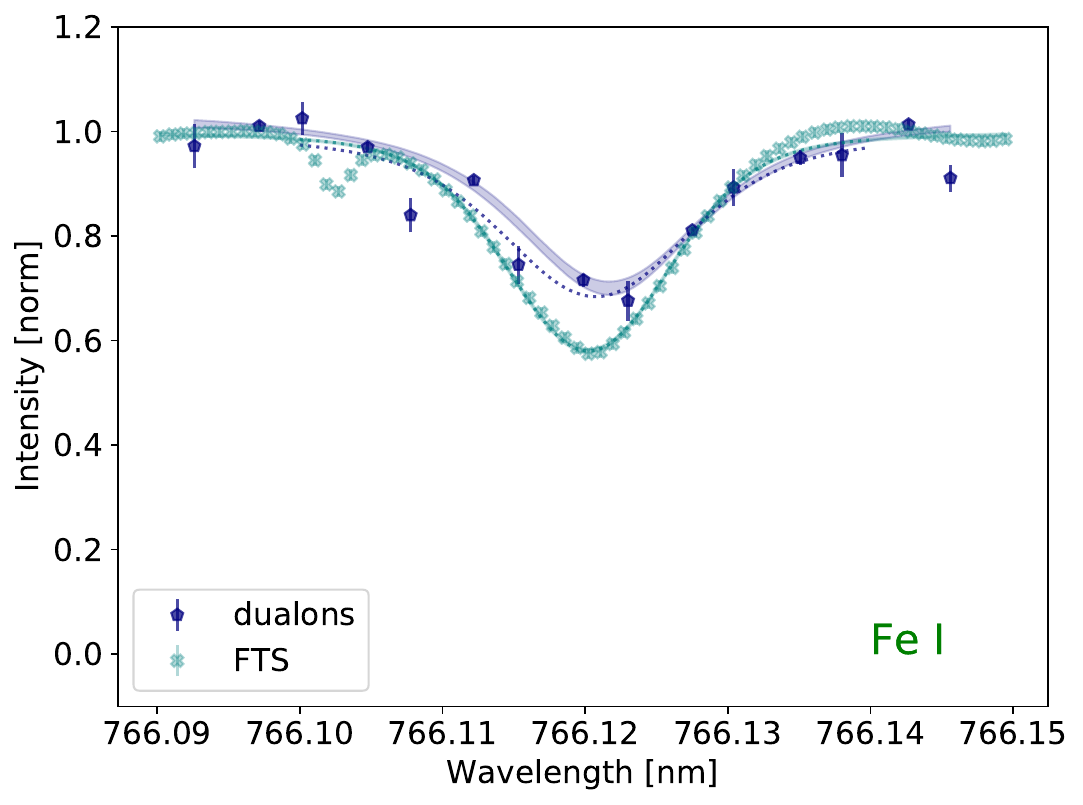} \\
\includegraphics[width=0.32\columnwidth]{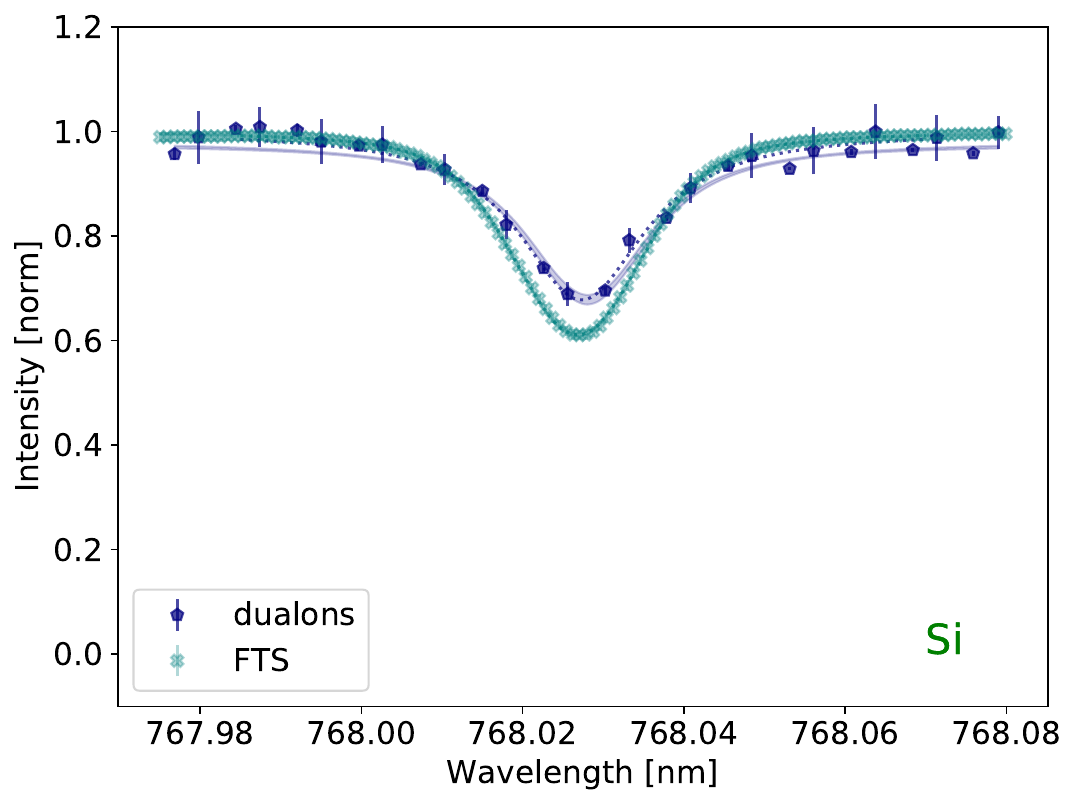} 
\includegraphics[width=0.32\columnwidth]{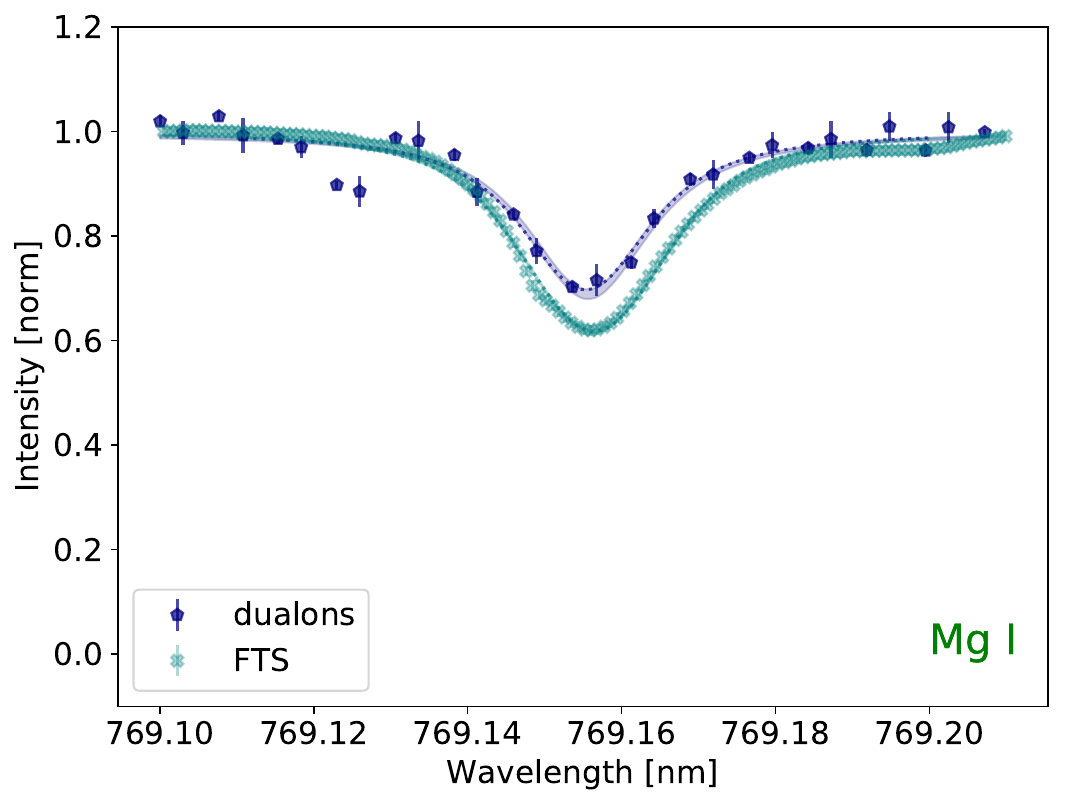} 
\includegraphics[width=0.32\columnwidth]{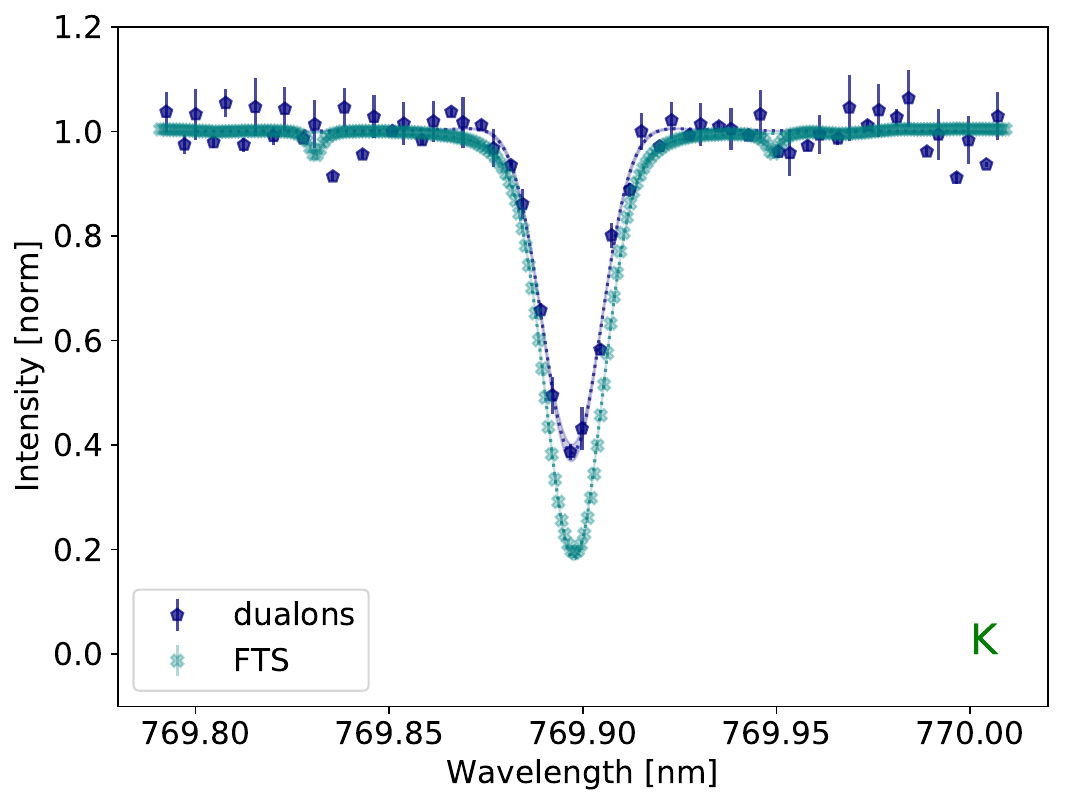} 
\end{center}
\caption{The zoomed region of some atomic lines detected with  FPI array (blue), and Fourier Transform Spectrograph (FTS) at Kitt Peak (green). The measurement detail is in Table \ref{tab:atomic lines} \label{fig:o2_atomic_lines}}
\end{figure}

After subtracting the telluric \mo, additional absorption lines are visible. Those features are marked in Figure \ref{fig:elements}. We mark known atomic absorption features using the NIST Atomic Spectra Database (ver. 5.9). Table~\ref{tab:atomic lines} listed the detected atomic lines within the \mo{}  region. Zoom-in regions for some of these lines are shown in Figure \ref{fig:o2_atomic_lines}, and compared with the FTS detection. It should be noted that the instrument resolution is higher than necessary to resolve these atomic lines.

\begin{table}[h]
\renewcommand{\thetable}{\arabic{table}}
\centering
\caption{Some atomic lines observed within $760-770\,\mathrm{nm}$ window resolved by FIOS and FTS. Unit are in nm.}  \label{tab:atomic lines}
\begin{tabular}{cccc}
\hline
$\lambda$ [nm] & Species & $\mathrm{\Delta\lambda_{dualon}}$& $\mathrm{\Delta\lambda_{FTS}}$ \\
\hline
\hline

761.921 & Ni I & 0.0154 $\pm$ 0.0005 & 0.0149 $\pm$ 0.0002\\ 
765.760 & Mg I & 0.0170 $\pm$ 0.0005 & 0.0196 $\pm$ 0.0002 \\
766.119 & Fe I &  0.0163 $\pm$ 0.0012 & 0.0138$\pm$ 0.0002 \\
766.429 & Fe I & 0.0226 $\pm$ 0.0002 & 0.0203 $\pm$ 0.0003\\
768.026 & Si & 0.0187 $\pm$ 0.0008 & 0.0193 $\pm$ 0.0002 \\
769.155 & Mg I & 0.0194 $\pm$ 0.0009 & 0.0228 $\pm$ 0.0003 \\
769.896 & K & 0.0178 $\pm$ 0.0003 & 0.0177 $\pm$ 0.0001 \\

\hline
\end{tabular}
\end{table}

\section{Discussion} \label{sec:discussion}

This work provides an end-to-end demonstration of an FPI array that can reach ultra-high resolution and exploit the FPI resolution. The results from these experiments suggest that the FPI array creates chain spectral profiles and reached high resolution depending on the fiber sizes in the same setup. This agrees well with the previous study in Rk20.  With a single-mode fiber, the instrument reached resolutions of R=700,000 (see Table \ref{tab:R_budget}). In this study we mainly used a 50 $\mu$m fiber. Once the instrumental profile is marginalized out with convolution, the system's resolution (FIOS and external spectrograph) is degraded to R=250,000. We expect that the results, taken with a 50 $\mu$m fiber, can be improved further to 300,000-400,000 by reducing the collimated beam divergence which results from the modest dualon aperture, see BA18.

By design, dualon works best at the position parallel to the incident angle. However, to achieve a sequence of arms where the reflected light can be captured, a small additional angle needed to be introduced. This beam deviation, and consequently operating the multi-dualon in a non-traditional position lowers the resolution compared with the original, theoretical expectations. In other words, an additional angle was introduced and caused the deviation of the beam and therefore lower resolution. This issue can be mitigated by a better ratio between the fiber size, collimator focal length and dualon aperture size. Our prototype was originally designed for a single mode fiber, $f_{col}$ = 30 mm and etalon aperture is 13 mm to reach a resolution of R=500,000. When the fiber size was altered, it affected the resolution of the instrument both the FPI array and the external spectrograph. However, in a previous study \cite{2017Cersullo} demonstrated that using a 200 $\mu$m fiber,  $f_{col}$ = 30 mm and etalon aperture is 40 mm for the same resolution. This implies that using a larger fiber input is possible, specifically, if we want to combine this FPI array with a spectrograph like G-CLEF or ANDES UBV spectrograph for the \mo{} A-band. Also we expect that for longer baseline exposures, thermal and pressure stabilization of the FPI array is a requirement.

We turn now to discuss losses from the external spectrograph. One option, discussed also in BA18 is to couple FIOS to a high-resolution spectrograph such as Espresso \citep{Pepe2014}, Maroon-X \citep{Seifahrt2022}, G-CLEF \citep{2016SPIE.9908E..22S} and ANDES \citep{Marconi2022}. These offer an end-to-end transmission efficiency of 5-10\%. While such coupling is feasible, it will result in an overall low efficiency that might make FIOS inexpedient. As FIOS is observing only a narrow bandpass of $\sim10\,$nm, a plausible solution was presented by \cite{Sofer_Rimalt2022} with an instrument that offers high resolution for $m=1$ interference order using optimized binary mask ion-etched gratings. Such an instrument has the potential of obtaining an efficiency of $\sim$70\%, and so we can postulate an overall efficiency of $\sim$30\% for FIOS when coupled to such a spectrograph. Alternatively, the VIPA spectrograph, with an estimated total transmission of 40\% \citep{Bourdarot2017,Zhu_2020} is a reasonably intermediate solution between the 5-10\% class and the 70\% transmission of \cite{Sofer_Rimalt2022}.

\begin{figure}[]
\begin{center}
\includegraphics[width=0.99\columnwidth]{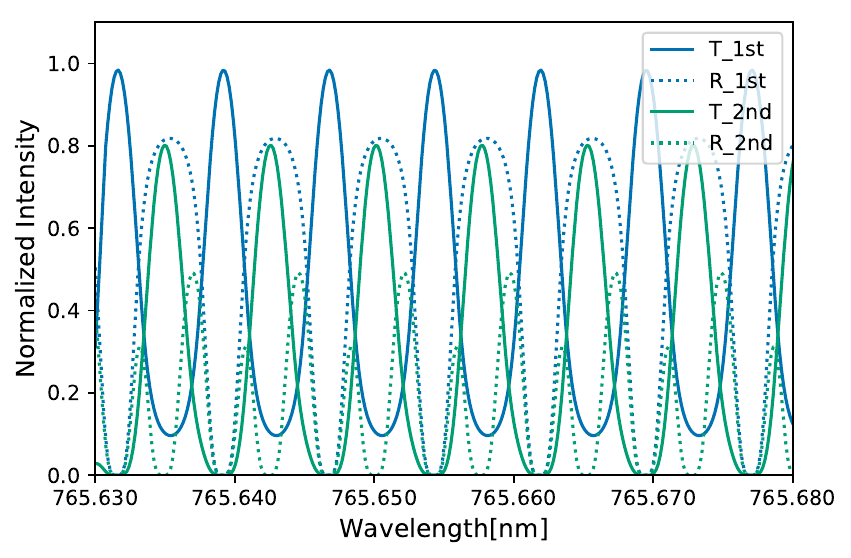} 
\caption{A model showing the signal level where an additional arm could be added into the FPI array with the same specification of the ones used in the prototype. The blue and green solid lines indicate the transmission ($I_T$) signal from the existing first and second arm. The dotted lines are the reflection ($I_R$) signal. \label{fig:2arms_convolved}}
\end{center}
\end{figure}

\begin{figure}[]
\begin{center}
\includegraphics[width=0.99\columnwidth]{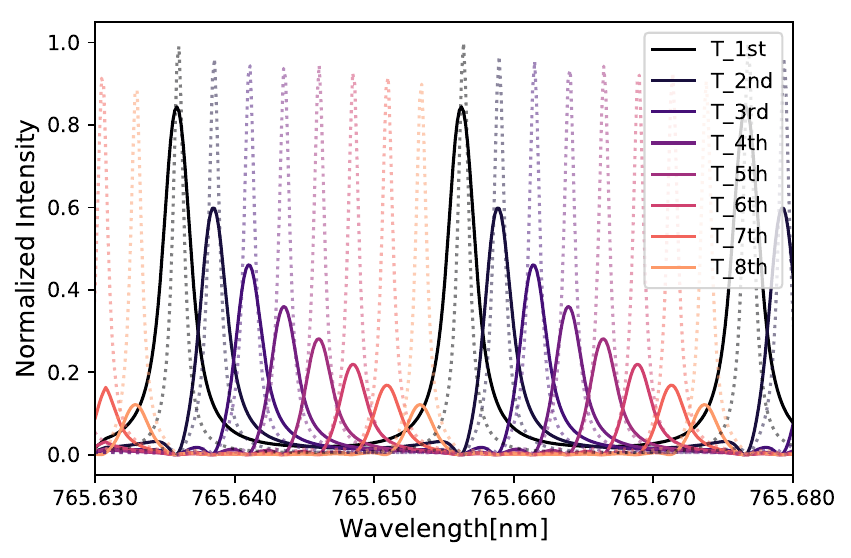} 
\caption{A realization prediction of the array of 8 FPI profiles chained together including the 20\% loss when an additional arm was introduced. The solid line is the signal with loss over-plotted on the dotted line of the theoretical model.   \label{fig:8arms_convolved}}
\end{center}
\end{figure}

For a future outlook, we incorporate the results obtained from this study to realistically predict the spectral profile of future, full FIOS instrument. To achieve the intended full chain (8 arms) design as described in BA18, modifications to our current system (2-arm FPI array and external spectrograph) are needed. We can infer from Figure~\ref{fig:2arms_convolved} that the current specification of the dualon used in this prototype cannot be extended to more dualons, e.g., to the 8-arm design as mentioned in BA18. To realize the 8-arm design, we assume a smaller FPI thickness of 9770 $\mu$m. The theoretical model has been revisited by taking into account the 20\% loss with signal convolution. The realistic signal prediction is displayed in Figure \ref{fig:8arms_convolved} in the solid line over-plotted on the theoretical model. The convolved model includes the separation (d) difference of 38 nm of each dualon. The incident angle was set to be 0.01 degrees apart from each arm starting with 0.08 degrees (5 arcmin). The fitted angle from the measurements in this study yields a maximum of 6.5 arcmin, which is a result of beam deviation \citep[previously experimented and discussed in][]{Rukdee_2020}.

The fundamental idea behind FIOS is that it can replace a single, huge ultra high resolution conventional spectrograph with a small FIOS unit and small low-to-medium resolution spectrograph. The main design of FIOS was based on the FPI (dualon) design, and the whole chain FIOS consists of 8 arm FPI. The chained profile is produced by incorporating the mismatch distance, which is a few nanometers, between each dualon. The optomechanical supports and other elements of the optical array are identical. The replication process of the several arms is indeed straightforward in terms of optical design and opto-mechanics, but not its integration/alignment as a full instrument, whose complexity (multi-arm instrument) is inherent to the concept presented here. In terms of sensitivity loss, each arm added introduces approximately 10\% loss from the previous arm. This instrument scheme is suitable for the external spectrograph implemented with the pupil slicer scheme under natural seeing conditions. It demonstrates the potential to boost the resolution of existing and planned R=100,000 spectrographs two to four-fold. 

It is challenging to go straight from a two-arm setup to an eight-arm configuration in a more complex system such as ELTs' instrument. Instead, we recommend an intermediate step by implementing a 4-arm array plug-in to an already-existing spectrograph on a medium-class telescope before incorporating the entire 8-arm FIOS into an ELT's spectrograph. This strategy provides the optimal solution for interim optimization.

Alternatively, once an adaptive optic (AO) system is activated (for second-generation Extremely Large Telescopes' instruments), it is also possible that a single-mode fiber-coupled spectrograph could reach an extremely high resolution, although the resolution could be limited by the Echelle resolution. We are also exploring AO options. Since an optical fiber spectrograph has never been fed with an AO system in the optical, further development is needed  \citep[see, e.g.,][]{Mello2018}.

%\newpage
\section{Conclusion} \label{sec:conclusion}

In this work, we experimentally demonstrate the ability of the two-arm prototype (1/4 of a full chain) of FIOS to record a telluric spectrum at a high resolution of R=250,000 and characterize narrow \mo{} spectral features under realistic conditions. We demonstrated how an additional arm helps improve the sampling frequency and overall resolution towards the upper R limit from the laser scan measurement. This result bodes well for the future development of future FIOS. The chained ultra-high resolution dualon array is cross-dispersed by a high-resolution spectrograph of similar spectral resolution as shown in Figure \ref{fig:ccdimages} bottom panel. By integrating over the transmission peaks resolved by the external spectrograph, we can emulate the behavior of a lower resolution spectrograph with a resolution similar to the dualons FSR, and showed this is sufficient to unlock an ultra-high resolution of the instrument and makes FIOS a plug-in resolution booster for a narrow band, while the external spectrograph minimum requirement is R = 100,000 (1 FSR of the FPI). 

The choice of multi-mode fiber was selected to achieve a reasonable throughput and aperture ratio among the optical components in the system, the external spectrograph on ELTs will use multi-mode large core size fiber under natural seeing conditions. This makes our demonstration realistic for the intended future of FIOS on ELTs. 

Beyond \mo{}, the concept can also be used to study other spectral bands/lines, for example, methane ($\rm CH_4$) where the lines are much denser than the oxygen band.  This concept can be used for the future exoplanet atmospheres study by plugging FIOS into upcoming instruments for the ELTs such as G-CLEF \citep{2016SPIE.9908E..22S}, ANDES \citep{2018SPIE10702E..1YM} or MODHIS \citep{2019BAAS_Mawet} with a minimal alteration from the original concept of these instruments. 

%\section*{Acknowledgements}
\begin{acknowledgements}
We would like to thank the referee for providing constructive comments and valuable inputs that have allowed us to enhance the clarity of our discussions. This work was made possible through the support of a grant from the John Templeton Foundation. The opinions expressed here are those of the authors and do not necessarily reflect the views of the John Templeton Foundation. We also thank the Brinson Foundation and the Smithsonian Institution for providing funding to support this project. 
\end{acknowledgements}

%\bibliography{ref}

\end{document}